\documentclass[showpacs,preprintnumbers,amsmath,amssymb,prd]{revtex4}
\usepackage{dcolumn}
\usepackage{bm}
\newcommand{\be}{\begin{equation}}
\newcommand{\ee}{\end{equation}}
\newcommand{\ba}{\begin{eqnarray}}
\newcommand{\ea}{\end{eqnarray}}
\newcommand{\nn}{\nonumber}
\def\etap{{\eta^{(')}}}

\begin{document}

\title{Nonleptonic two-body $B$-decays including axial-vector mesons 
in the final state}

\author{G. Calder\'on}
\email{gecalde@gmail.com}
\affiliation{Facultad de Ingenier\'ia Mec\'anica y El\'ectrica,
Universidad Aut\'onoma de Coahuila, C.P. 27000, Torre\'on, Coahuila, M\'exico}
\author{J. H. Mu\~noz}
\email{jhmunoz@ut.edu.co}
\author{C. E. Vera}
\email{cvera@ut.edu.co}
\affiliation{Departamento de F\'isica, Universidad del Tolima A. A.
546, Ibagu\'e, Colombia}

\begin{abstract}
We present a systematic study of exclusive charmless nonleptonic two-body $B$ decays 
including axial-vector mesons in the final state. We calculate branching ratios of 
$B\to PA$, $VA$ and $AA$ decays, where $A$, $V$ and $P$ denote an axial-vector, a vector 
and a pseudoscalar meson, respectively. We assume naive factorization hypothesis and use 
the improved version of the nonrelativistic ISGW quark model for form factors in $B\to A$ 
transitions. We include contributions that arise from the effective $\Delta B=1$ weak
Hamiltonian $H_{eff}$. The respective factorized amplitude of these decays are explicitly 
showed and their penguin contributions are classified. We find that decays 
$B^-\to a_1^0\pi^-$, $\bar B^0\to a_1^{\pm}\pi^{\mp}$, $B^-\to a_1^-\bar K^{0}$, 
$\bar B^0\to a_1^+K^-$, $\bar B^0\to f_1\bar K^0$, $B^-\to f_1K^-$, $B^-\to K_1^-(1400)\etap$, 
$B^-\to b_1^-\bar K^{0}$, and $\bar B^0\to b_1^+\pi^-(K^-)$ have branching ratios of the order 
of $10^{-5}$. We also study the dependence of branching ratios for $B \to K_1P(V,A)$ decays 
($K_1=K_1(1270),K_1(1400)$) with respect to the mixing angle between $K_A$ and $K_B$.
\end{abstract}


\maketitle

\section{Introduction}

Recent experimental results for $B\to a_1\pi$ and $B\to K_1(1270)\gamma$ decays
obtained by BABAR, Belle and CLEO \cite{exp} have opened an interesting area of research 
about production of axial-vector mesons in $B$ decays. Two-body $B$ decays have been 
considered one of the premier places to understand the interplay of QCD and electroweak 
interactions, to look for CP violation and over constrain the CKM parameters in the 
Standard Model. And indeed, exclusive modes $B\to PP$, $PV$ and $VV$, which have been 
extensively discussed in the literature have committed such expectations.

In the search of different alternative modes to the traditional studied, we consider
processes which include an axial-vector meson in the final state. It is expected that some 
of these decay channels have large branching ratios \cite{nardulli05} and can be within the 
reach of future experiments. Moreover, they are an additional scenario for understanding QCD 
and electroweak penguin effects in the Standard Model. These modes give additional and 
complementary information about exclusive nonleptonic weak decays of $B$ mesons.

The two most important penguin contributions correspond to $a_4$ and $a_6$ QCD coefficients.
These coefficients have different sign in the amplitude ${\cal M}(B\to VP)$, making their
contribution small. For $B\to AP$ decays they have equal sign, thus we have a bigger
contribution in penguin sector. Branching ratios of these decays are good candidates to
be measured. 

Our purpose is to present a systematic analysis about charmless modes $B\to AP$, $B\to AV$
and $B\to AA$, similar in completeness to previous studies about channels $B\to PP$, $B\to PV$ 
and $B\to VV$ which have been extensively considered in the literature \cite{ali,chen99}.

There are two types of axial-vector mesons \cite{pdg2006}. In spectroscopic notation
$^{2s+1}L_J$, these $p$-wave mesons are $^3P_1$ and $^1P_1$, with $J^{PC} =1^{++}$ and $1^{+-}$,
respectively. Under $SU(3)$ flavor symmetry, the $^3P_1$-nonet is composed by $a_1(1260)$,
$f_1(1285)$, $f_1(1420)$, and $K_{1A}$ and the $^1P_1$-nonet is integrated by $b_1(1235)$,
$h_1(1170)$, $h_1(1380)$, and $K_{1B}$. However, physical strange axial-vector mesons
$K_1(1270)$ and $K_1(1400)$ are a mixture of $K_{1A}$ and $K_{1B}$

\ba K_1(1270) &=& K_{1A}\sin \theta + K_{1B}\cos \theta\nn\\
K_1(1400) &=& K_{1A}\cos \theta - K_{1B}\sin \theta\ , \label{Kmatrix} \ea

\noindent where $\theta$ is the mixing angle.

At theoretical level some authors have worked with production of axial-vector mesons in
nonleptonic $B$ decays. Katoch-Verma \cite{kave96} studied $B\to PA$ decays at tree level
using the factorization hypothesis and the non-relativistic Isgur-Scora-Grinstein-Wise (ISGW)
quark model \cite{isgw89}. Nardulli-Pham in Ref. \cite{nardulli05} did an analysis of two-body
$B$ decays with an axial-vector meson in final state using factorization and the $B\to K_1$
form factors obtained from measured radiative decays. They calculated the branching ratio
for $B\to J/\psi K_1$ and derived some predictions for a few nonleptonic decays channels
involving light strange or non-strange axial-vector mesons in final state using naive
factorization and relations from Heavy Quark Effective Theory. Recently, Laporta-Nardulli-Pham
\cite{nardulli06} presented an analysis about some charmless $B\to PA$ decays including
contributions of the effective weak Hamiltonian $H_{eff}$, assuming factorization approach and
employing as inputs a limited number of experimental data. They did not use predictions from
theoretical models for form factors. In Ref. \cite{chen05}, the authors investigated
$B\to K_1\phi$ decays employing the generalized factorization hypothesis and light-front
approach for form factors.

Cheng in Ref. \cite{cheng03} studied Cabibbo-allowed hadronic $B$ decays at tree level
containing an even-parity charmed meson in final state. In this work, the author
predicted branching ratios for some decays of type $B\to AP(V)$ where $A$ is,
in this case, a charmed axial-vector meson. Calculation was performed within the
framework of generalized factorization. Form factors for $B\to A$ transition were
calculated with the improved version of the Isgur-Scora-Grinstein-Wise quark model,
called ISGW2 \cite{isgw95}. For $B\to P$ and $B\to V$ form factors, the author used the
Melikhov-Stech model \cite{msmodel}. Recently, Cheng-Chua \cite{cheng06} continue
studying even-parity charmed meson production in $B$ decays, calculating $B\to D^{**}$
($D^{**}$ denotes a $p$-wave charmed meson) form factors within the covariant light-front
quark model.

Others authors have been interested in radiative $B\to K_1$ decays (see for example Ref.
\cite{nardulli05}). Recently, J-P. Lee \cite{lee06} revisited the $B\to K_1$ form factors
in the light-cone sum rules, and reduced the discrepancy between theoretical prediction
and experimental data reported by Belle Collaboration \cite{exp} for $B\to K_1\gamma$. Lee
claims that it is necessary more information about the mixing angle between $K_{1A}$ and
$K_{1B}$ to reduce theoretical uncertainties. In fact, this mixing angle has been estimated
by some different methods \cite{mixing}. However there is not yet a consensus about its
value \cite{lili06}.

CP violation effects have been also investigated in nonleptonic $B$ decays with axial-vector
mesons in the final state. For example, in Ref. \cite{gronau03} time-dependent CP asymmetries
in $\bar B^0\to D^{*-}a_1^+$ are studied in order to learn about the linear combination of
weak phases $(2\beta+\gamma)$. And more recently (see Ref. \cite{gronau06}), in an analysis
of $B^0\to a^{\pm}_1 \pi^{\mp}$ modes, they determined the phase $\alpha_{eff}$, which include
the weak phase $\alpha$ and effects due to penguin contribution. Moreover, applying $SU(3)$
symmetry to these decays and to $B\to a_1 K$ and $B\to K_1\pi$, they obtained bounds on
($\alpha-\alpha_{eff}$).

In this paper we are interested in studying exclusive charmless nonleptonic two-body
$B$ decays including axial-vector mesons in the final state. We present an overview and a
systematic study about this type of processes. For this, we compute branching ratios
for exclusive channels $B\to AP$, $AV$, $AA$ that are allowed by the CKM factors,
including contributions of the effective weak Hamiltonian $H_{eff}$ (tree and penguin),
assuming the naive factorization hypothesis and using the improved version of the ISGW
\cite{isgw95} quark model for calculating the respective form factors related with
$B\to A$ transitions. Form factors for $B\to P$ and $B\to V$ transitions have been taken
from the relativistic Wirbel-Stech-Bauer (WSB) quark model \cite{wsb} and from Light
Cone Sum Rules (LCSR) \cite{ball}.

This paper is organized as follows: In Sec. II we discuss the effective weak Hamiltonian,
effective Wilson coefficients and naive factorization hypothesis. Input parameters and
mixing schemes are discussed in Sec. III. In Sec. IV we present form factors for $B\to P(V)$ 
transitions taken from the WSB model and LCSR approach, and $B\to A$ transitions calculated 
in ISGW2 model. In Sec. V is discussed the amplitudes and manner to calculate branching ratios 
for processes considered. Numerical results for branching ratios are presented in Sec. VI. 
We conclude in Sec. VII with a summary. Amplitudes for all charmless $B\to AP$, $AV$ and 
$AA$ processes are given explicitly in the appendices.

\section{Effective Hamiltonian and factorization}

The basis for the study of two-body charmless hadronic $B$-decays is the effective weak
Hamiltonian $H_{eff}$ \cite{buras96}. For $\Delta B=1$ transitions it can be written as

\ba H_{eff} &=& \frac{G_F}{\sqrt{2}}\Bigg[V_{ub}V^*_{uq}
\Big(C_1(\mu) O^u_1(\mu) + C_2(\mu) O^u_2(\mu)\Big)
+ V_{cb}V^*_{cq}\Big(C_1(\mu) O^c_1(\mu) + C_2(\mu) O^c_2(\mu)\Big)\nn\\
&&- V_{tb}V^*_{tq}\Bigg(\sum^{10}_{i=3}C_i(\mu) O_i(\mu)
+ C_g(\mu) O_g(\mu)\Bigg) \Bigg] + h.c.\ ,\ea

\noindent where $G_F$ is the Fermi constant and $C_i(\mu)$ are the Wilson coefficients
evaluated at the renormalization scale $\mu$. Local operators $O_i(\mu)$ are given below 
for $b\to q$ transitions

\ba O^u_1 &=& \bar q_\alpha\gamma^\mu L u_\alpha \cdot
\bar u_\beta \gamma_\mu L b_\beta \nn\\
O^u_2 &=& \bar q_\alpha\gamma^\mu L u_\beta \cdot
\bar u_\beta \gamma_\mu L b_\alpha \nn\\
O^c_1 &=& \bar q_\alpha\gamma^\mu L c_\alpha \cdot
\bar c_\beta \gamma_\mu L b_\beta \nn\\
O^c_2 &=& \bar q_\alpha\gamma^\mu L c_\beta \cdot
\bar c_\beta \gamma_\mu L b_\alpha \nn\\
O_{3(5)} &=& \bar q_\alpha\gamma^\mu L b_\alpha \cdot
\sum_{q'} \bar q'_\beta \gamma_\mu L(R) q'_\beta \nn\\
O_{4(6)} &=& \bar q_\alpha\gamma^\mu L b_\beta \cdot
\sum_{q'} \bar q'_\beta \gamma_\mu L(R) q'_\alpha \nn\\
O_{7(9)} &=& \frac{3}{2}\bar q_\alpha\gamma^\mu L b_\alpha \cdot
\sum_{q'} e_{q'}\bar q'_\beta \gamma_\mu R(L) q'_\beta \nn\\
O_{8(10)} &=& \frac{3}{2}\bar q_\alpha\gamma^\mu L b_\beta \cdot
\sum_{q'} e_{q'}\bar q'_\beta \gamma_\mu R(L) q'_\alpha\nn\\
O_g&=&(g_s/8\pi^2)m_b\bar q_\alpha\sigma^{\mu\nu}
R(\lambda_{\alpha\beta}^A/2)b_{\beta}G^A_{\mu\nu}\ ,\ea

\noindent where $q$ can be the quarks $d$ or $s$. $L$ and $R$ stand for left and right
projectors defined as $(1-\gamma_5)$ and $(1+\gamma_5)$, respectively. The symbols
$\alpha$ and $\beta$ are $SU(3)$ color indices and $\lambda_{\alpha\beta}^A$ ($A=1, ...,8$)
are the Gell-Mann matrices. The sums run over active quarks at the scale
$\mu={\cal O}(m_b)$, i.e. $q'$ runs with the quarks $u$, $d$, $s$ and $c$.

We use the next to leading order Wilson coefficients for $\Delta B=1$ transitions
obtained in the naive dimensional regularization scheme (NDR) at the energy scale
$\mu=m_b(m_b)$, $\Lambda^{(5)}_{\overline{MS}}=225$ MeV and $m_t=170$ GeV. These values are
$c_1=1.082$, $c_2=-0.185$, $c_3=0.014$, $c_4=-0.035$, $c_5=0.009$, $c_6=-0.041$,
$c_7/\alpha=-0.002$, $c_8/\alpha=0.054$, $c_9/\alpha=-1.292$ and $c_{10}/\alpha=0.263$,
where $\alpha$ is the fine structure constant, see Table XXII in Ref. \cite{buras96}.

In order to calculate the amplitude for a nonleptonic two-body $B\to M_1 M_2$ decay,
we use the effective weak Hamiltonian $H_{eff}$,

\ba {\cal M}(B\to M_1 M_2) &=& \langle M_1 M_2|H_{eff}|B\rangle
=\frac{G_F}{\sqrt{2}}\sum_i C_i(\mu)\langle M_1 M_2|O_i(\mu)|B\rangle\ .\ea

Hadronic matrix elements
$\langle O_i(\mu)\rangle\equiv \langle M_1 M_2|O_i(\mu)|B\rangle$ can be
evaluated under factorization hypothesis, which approximates hadronic matrix
element by product of two matrix elements of singlet currents. These currents are
parametrized by decay constants and form factors. It is necessary to point out that
these matrix elements as products of conserved currents are $\Lambda_{\overline{MS}}$,
$\mu$ scale and renormalization scheme independent \cite{buras95, buras98}. The
suggested energy scale to apply factorization for $B$ decays is $\mu_f={\cal O}(m_b)$.
Besides this simple approximation, it is well established that nonfactorisable
contributions must be present in the matrix elements in order to cancel the scale $\mu$
and renormalization scheme dependence of $C_i(\mu)$.

To solve the issue of scale $\mu$ dependence, but not the renormalization scheme
dependence \cite{buras99}, it is proposed in Refs. \cite{ali,chen99}, to isolate from
the matrix element $\langle O_i(\mu)\rangle$ the $\mu$ dependence, and link  with the
$\mu$ dependence in the Wilson coefficients $C_i(\mu)$ to form $c^{eff}_i$, effective
Wilson coefficients independent of $\mu$. Matrix elements $\langle O_i\rangle_{tree}$
and effective $c^{eff}_i$ Wilson coefficients are scale $\mu$ independent, so the
amplitude. Thus, we can write

\ba \sum_i C_i(\mu)\langle O_i(\mu)\rangle =
\sum_i C_i(\mu)g_i(\mu)\langle O_i\rangle_{tree} =
\sum_i c^{eff}_i\langle O_i\rangle_{tree}\ .\ea

The formula for effective Wilson coefficients and their numerical values have been given
explicitly in Ref. \cite{chen99}. These values depend on quark masses, CKM parameters and
renormalization scheme. In this article we recalculate the effective Wilson coefficient
$c^{eff}_i$, because there have been some changes in the CKM parameters since the authors
of Ref. \cite{chen99} calculated them. We choose naive dimensional regularization scheme 
to calculate. We present the effective Wilson coefficients $c^{eff}_i$ in Table I,
for $b\to d$ and $b\to s$ transitions evaluated at the factorization scale $\mu_f=m_b$,
averaged momentum transfer $k^2=m^2_b/2$ and current CKM parameters, see Sec. III.

The effective Wilson coefficients appear in decay amplitudes as linear combinations.
This allows to define $a_i$ coefficients, which encode dynamics of the decay, by

\ba a_i &\equiv& c^{eff}_i + \frac{1}{N_c} c^{eff}_{i+1}\ (i=odd)\nn\\
a_i &\equiv& c^{eff}_i + \frac{1}{N_c} c^{eff}_{i-1}\ (i=even)\ ,\ea

\noindent where index $i=1,...,10$ and $N_c=3$ is the color number. In Table II, we give 
the $a_i$ values for $b\to d$ and $b\to s$ transitions calculated with the effective Wilson
coefficients $c^{eff}_i$, given in Table I.

\begin{table}[ht]
{\small TABLE I.~Effective Wilson coefficients $c^{eff}_i$ for $b\to d$ and $b\to s$
transitions. Evaluated at $\mu_f=m_b$ and $k^2=m^2_b/2$, where we use the Wolfenstein
parameters $\lambda=0.2272$, $A=0.818$, $\rho=0.227$ and $\eta=0.349$, see Sec. III.}
\begin{center}
\begin{tabular}{l c c}
\hline \hline
$c^{eff}_i$           & $b\to d$           & $b\to s$ \\
\hline
$c^{eff}_1$           &  1.1680            &  1.1680           \\
$c^{eff}_2$           & -0.3652            & -0.3652            \\
$c^{eff}_3$           &  0.0233 + i 0.0036 &  0.0233 + i 0.0043 \\
$c^{eff}_4$           & -0.0481 - i 0.0109 & -0.0482 - i 0.0129 \\
$c^{eff}_5$           &  0.0140 + i 0.0036 &  0.0140 + i 0.0043 \\
$c^{eff}_6$           & -0.0503 - i 0.0109 & -0.0504 - i 0.0129 \\
$c^{eff}_7/\alpha$    & -0.0310 - i 0.0317 & -0.0312 - i 0.0357 \\
$c^{eff}_8/\alpha$    &  0.0551            &  0.0551            \\
$c^{eff}_9/\alpha$    & -1.4275 - i 0.0317 & -1.4277 - i 0.0357 \\
$c^{eff}_{10}/\alpha$ &  0.4804            &  0.4804            \\
\hline \hline
\end{tabular}
\end{center}
\end{table}

\begin{table}[ht]
{\small TABLE II.~Effective coefficients $a_i$ for $b\to d$ and $b\to s$ transitions
(in unit of $10^{-4}$ for $a_3$, ..., $a_{10}$).}
\begin{center}
\begin{tabular}{l c c}
\hline \hline
$a_i$    & $b\to d$        &  $b\to s$       \\
\hline
$a_1$    & 1.046           &  1.046          \\
$a_2$    & 0.024           &  0.024          \\
$a_3$    & 72              &  72             \\
$a_4$    & -403   - i 97   & -404   - i 115  \\
$a_5$    & -28             & -28             \\
$a_6$    & -456   - i 97   & -457   - i 115  \\
$a_7$    & -0.92  - i 2.31 & -0.94  - i 2.61 \\
$a_8$    &  3.26  - i 0.77 &  3.26  - i 0.87 \\
$a_9$    & -92.5  - i 2.31 & -92.5  - i 2.61 \\
$a_{10}$ &  0.33  - i 0.77 &  0.33  - i 0.87 \\
\hline \hline
\end{tabular}
\end{center}
\end{table}

\section{Input parameters}

We parametrize the CKM matrix in terms of the Wolfenstein parameters
$\lambda$, $A$, $\bar\rho$ and $\bar\eta$ \cite{wolfenstein}

\ba \left(\begin{array}{ccc}
1-\frac{1}{2}\lambda^2 & \lambda & A\lambda^3(\rho-i\eta) \\
-\lambda & 1-\frac{1}{2}\lambda^2 & A\lambda^2 \\
A\lambda^3(1-\bar\rho-i\bar\eta) & -A\lambda^2 & 1
\end{array}\right) \ea

\noindent with $\bar\rho=\rho(1-\lambda^2/2)$ and $\bar\eta=\eta(1-\lambda^2/2)$,
including ${\cal O}(\lambda^5)$ corrections \cite{buras94}.

By a global fit that uses all available measurements and that imposes unitary constrains,
the Wolfenstein parameters are precisely determined. There exist two ways to combining
experimental data, the frequentist statistics \cite{ckmfitter} and the Bayesian approach
\cite{utfit}, providing similar results. Thus, we take for the Wolfenstein parameters the
central values $\lambda=0.2272$, $A=0.818$, $\bar\rho=0.221$ and $\bar\eta=0.340$
\cite{pdg2006}.

The running quark masses are necessary in calculation of penguin terms in the amplitude
where appear scalar and pseudoscalar matrix elements which are reduced by use of Dirac
equation of motion. Running quark masses are given at the scale $\mu \approx m_b$, since
energy released in $B$ decays is of order $m_b$. We use $m_u(m_b)=3.2$ MeV,
$m_d(m_b)=6.4$ MeV, $m_s(m_b)=127$ MeV, $m_c(m_b)=0.95$ GeV and $m_b(m_b)=4.34$ GeV, see
Ref. \cite{fusaoku}.

Decay constants of pseudoscalar and vector mesons are well determined experimentally.
We use the following values \cite{pdg2006}: $f_\pi=130.7$ MeV, $f_K=160$ MeV,
$f_\rho=216$ MeV, $f_\omega=195$ MeV, $f_{K^\star}=221$ MeV and $f_\phi=237$ MeV.

The $\omega-\phi$, $\rho^0-\omega$, $\eta-\eta'$ and $K_{1A}-K_{1B}$ mixing are introduced
through mixing in decay constants and form factors. We consider ideal mixing for the
system $(\omega, \phi)$, i.e. $\omega=1/\sqrt{2}(u\bar u+d\bar d)$ and $\phi=s\bar s$.
In next section we will discuss mixing in form factors. In the following we describe
mixing in decay constants.

For the $\eta-\eta'$ mixing we use the two mixing angle formalism proposed in
\cite{leutwyler, feldmann}, which define physical states $\eta$ and $\eta'$ in terms
of flavor octet and singlet, $\eta_8$ and $\eta_0$, respectively:

\ba && |\eta \rangle = \cos \theta_8 |\eta_8\rangle -
\sin \theta_0 |\eta_0\rangle\ ,\nn\\
&& |\eta' \rangle = \sin \theta_8 |\eta_8\rangle +
\cos \theta_0 |\eta_0\rangle\ .\ea

We introduce decay constants for $\eta_8$ and $\eta_0$ by
$\langle 0|A^8_\mu|\etap(p)\rangle = i f^8_\etap p_\mu$ and
$\langle 0|A^0_\mu|\etap(p)\rangle = i f^0_\etap p_\mu$.
Considering that $\eta_8$ and $\eta_0$ in terms of quarks are

\ba &&|\eta_8 \rangle = \frac{1}{\sqrt{6}}
|\bar u u + \bar d d - 2\bar s s\rangle\ ,\nn\\
&& |\eta_0 \rangle = \frac{1}{\sqrt{3}} |\bar u u + \bar d d + \bar
s s\rangle\ ,\ea

\noindent induce a two-mixing angle in decay constants $f^q_{\etap}$, defined by
$\langle 0|\bar q \gamma_\mu \gamma_5 q| \etap(p)\rangle = i f^q_{\etap}p_\mu$,

\ba && f^u_{\eta'} = \frac{f_8}{\sqrt{6}}\sin \theta_8 +
\frac{f_0}{\sqrt{3}}\cos \theta_0\ ,\nn\\
&& f^s_{\eta'} = -2\frac{f_8}{\sqrt{6}}\sin \theta_8 +
\frac{f_0}{\sqrt{3}}\cos \theta_0 \label{dc1}\ea

\noindent and

\ba && f^u_{\eta} = \frac{f_8}{\sqrt{6}}\cos \theta_8 -
\frac{f_0}{\sqrt{3}}\sin \theta_0\ ,\nn\\
&& f^s_{\eta} = -2\frac{f_8}{\sqrt{6}}\cos \theta_8 -
\frac{f_0}{\sqrt{3}}\cos \theta_0\ . \label{dc2}\ea

From a complete phenomenological fit of the $\eta-\eta'$ mixing parameters in Ref.
\cite{feldmann} we have $\theta_8=-21.1^\circ$, $\theta_0=-9.2^\circ$, $\theta=-15.4^\circ$,
$f_8=165$ MeV and $f_0=153$ MeV. Replacing values in Eqs. (\ref{dc1},\ref{dc2}), the
decay constants are $f^u_{\eta'}=61.8$ MeV, $f^s_{\eta'}=138$ MeV, $f^u_\eta=76.2$ MeV and
$f^s_\eta=-110.5$ MeV. To include in the mixing scheme the $\eta_c$, in the calculation we
use decay constants defined by
$\langle 0|\bar c \gamma_\mu \gamma_5 c |\etap \rangle=i f^q_{\etap}p_\mu$ as are obtained
in Ref. \cite{feldmann}: $f^c_\eta = -(2.4\pm 0.2)$ MeV and $f^c_{\eta'}=-(6.3\pm 0.6)$ MeV.

In evaluating hadron matrix elements of scalar and pseudoscalar densities in some penguin
terms  the anomaly must be included in order to ensure a correct chiral behavior for that 
matrix elements. The expressions are \cite{cheng98}

\ba && \langle \etap|\bar u\gamma_5 u|0\rangle =
\langle \etap|\bar d\gamma_5 d|0\rangle =
r_{\etap} \langle \etap|\bar s\gamma_5 s|0\rangle\ ,\nn\\
&& \langle \etap|\bar s\gamma_5 s|0\rangle =
-i\frac{m^2_\etap}{2m_s}\Big(f^s_\etap - f^u_\etap \Big)\ ,\ea

\noindent where the ratios $r_{\eta'}$ and $r_\eta$ are defined by

\ba && r_{\eta'} = \frac{\sqrt{2f^2_0-f_8}}{\sqrt{2f^2_8-f^2_0}}
\, \frac{\cos \theta + (1/\sqrt{2})\sin \theta}
{\cos \theta - \sqrt{2}\sin \theta}\ ,\nn\\
&& r_\eta = -\frac{1}{2}\frac{\sqrt{2f^2_0-f_8}}{\sqrt{2f^2_8-f^2_0}}
\, \frac{\cos \theta - \sqrt{2}\sin \theta}
{\cos \theta + (1/\sqrt{2})\sin \theta}\ ,\ea

\noindent the numerical values obtained are $r_{\eta'}=0.462$ and $r_\eta=-0.689$.

The physical states $K_1(1270)$ and $K_1(1400)$ result from the mixing of $K_{1A}$
and $K_{1B}$, $^3 P_1$ and $^1 P_1$ mesons, respectively, see Eq. (\ref{Kmatrix}). From
experimental data on masses and partial ratios of $K_1(1270)$ and $K_1(1400)$, it is
found two solutions for the mixing angle with a two-fold ambiguity, $\theta=\pm 32^\circ$
and $\theta=\pm 58^\circ$. The masses for the states $K_{1A}$ and $K_{1B}$ are
$m_{K_{1A}}=1367$ MeV and $m_{K_{1B}}=1310$ MeV, respectively. From $\tau$ decays, the decay
constants of the physical states are determined. The values obtained are $f_{K1}(1270)=171$
MeV and $f_{K1}(1400)=126$ MeV \cite{nardulli05}, using data from Ref. \cite{pdg2006}.

Thus, we have experimental information to determine decay constants for strange
axial-vector mesons. That is no the case for non-strange axial-vector mesons. But, using 
the mixing angle of the system $K_A-K_B$ and $SU(3)$ symmetry it is derived decay constants 
for axial-vector mesons: $(f_{b_1},f_{a_1})=(74,215)$ MeV for $\theta=32^\circ$ and 
$(f_{b_1},f_{a_1})=(-28,223)$ MeV for $\theta=58^\circ$, see Ref. \cite{nardulli05}. Since, 
$f_1$ and $h_1$ are in the same nonet that $a_1$ and $b_1$, respectively, by $SU(3)$ 
symmetry we consider equal decay constants. In the calculations of branching ratios we use 
the values $f_{a_1}=f_{f_1}=215$ MeV and $f_{b_1}=f_{h_1}=74$ MeV. However, in the exact 
limit of $SU(3)$ symmetry we have $f_{b_1}=f_{h_1}=0$. We calculate branching ratios and 
determine the modes which change in that limit.

Matrix elements for $B\to K_1$ transitions are calculated in the flavor base $K_{1A}-K_{1B}$.
In calculation of amplitude involving a final physical state as $K_1(1270)$ or $K_1(1400)$
we transform matrix elements from the flavor base to the physical base using
Eq. (\ref{Kmatrix}).

We use for $B$ meson lifetime $\tau_{B^-}=(1.638\pm 0.011)\times 10^{-12} s$ and
$\tau_{B^0}=(1.530\pm 0.009)\times 10^{-12} s$, see Ref. \cite{pdg2006}, necessary to 
calculate branching ratios.

\section{Form factors}

As was stated in Sec. II, hadronic matrix elements $\langle O_i\rangle_{tree}$ are given
in the factorization hypothesis in terms of decay constants and form factors.
Unfortunately, due to nonperturbative nature of these matrix elements, there is no
complete reliable calculations and only model dependent evaluations are used for them.

We use the WSB model and LCSR approach to determine form factors for $B\to P$ and $B\to V$ 
transitions. In the WSB model and LCSR approach, the form factors for $B\to A$ transitions 
have not been calculated. Thus we calculate form factors for $B\to A$ transitions in the 
ISGW2 model \cite{isgw95}. In the following subsections we give relevant information to 
calculate form factors in the respective models.

\subsection{Form factors for $B\to P(V)$ in the WSB model and LCSR approach}

In the WSB quark model meson-meson matrix elements of currents are evaluated from the
overlap integrals of corresponding wave functions, which are solutions of a
relativistic harmonic oscillator potential. For momentum transfer squared $q^2$
dependence of form factors in the region where $q^2$ is not too large, we shall use a 
single pole dominance ansatz, namely

\be f(q^2)=\,\frac{f(0)}{\left(1-{q^2/m^2_*}\right)}\ ,\ee

\noindent where $m_*$ is the pole mass and $f(0)$ the form factor at zero momentum
transfer given in Ref. \cite{wsb}. Note that the original WSB quark model assumes a 
monopole behavior for all form factors.

The WSB model has been quite successful in accommodating data in an important number of
exclusive semileptonic and nonleptonic two-body decays of $D$ and $B$ mesons.

Form factors for $B\to P$ transitions are defined as follows

\ba \langle P(p_P)|V_\mu| B(p_B)\rangle &\equiv& \Bigg[(p_B+p_P)_\mu
-\frac{m^2_B-m^2_P}{q^2}\ q_\mu\Bigg]F_1(q^2)
+\Bigg[\frac{m^2_B-m^2_P}{q^2}\Bigg]q_\mu F_0(q^2)\ ,\ea

\noindent where $q=(p_B-p_P)$, as well as form factors for $B\to V$ transitions by

\ba \langle V(p_V,\epsilon)|(V_\mu-A_\mu)| B(p_B)\rangle &\equiv&
-\epsilon_{\mu \nu \alpha \beta}\epsilon^{\nu *}p^\alpha_B p^\beta_V
\frac{2V(q'^2)}{(m_B+m_V)}-i\Bigg[\left(\epsilon^*_\mu-
\frac{\epsilon^*\cdot q'}{q'^2}q'_\mu\right)(m_B+m_V)A_1(q^2)\\
&&-\left((p_B+p_V)_\mu-\frac{(m^2_B-m^2_V)}{q'^2}q'_\mu\right)(\epsilon^*\cdot
q') \frac{A_2(q'^2)}{(m_B+m_V)} + \frac{2m_V(\epsilon^*\cdot
q')}{q'^2}q_\mu A_0(q'^2) \Bigg]\ ,\label{ffvector}\nn\ea

\noindent where $q'=(p_B-p_V)$ and $\epsilon$ is the polarization vector of $V$. In order
to cancel the poles at $q^2=0$, we must impose restrictions over form factors

\ba F_1(0) &=& F_0(0)\ ,\nn\\
2m_V A_0(0) &=& (m_B+m_V)A_1(0)-(m_B-m_V)A_2(0)\ .\ea

In Table III form factors are given for transitions required in calculations: form 
factors for $B\to \pi$, $B\to K$, $B\to \eta$, $B\to \eta'$, $B\to \rho$, $B\to K^*$ and 
$B\to \omega$ are evaluated at the $q^2=0$ momentum transfer. With respect to $B\to \eta$ 
and $B\to \eta'$ transitions the WSB model does not include the $\eta-\eta'$ mixing effect. 
We better consider $SU(3)$ symmetry and use the relations
$F^{B\pi}_0(0)=\sqrt{3}F^{B\eta_0}(0)=\sqrt{6}F^{B\eta_8}(0)$,
calculating physical form factors from

\ba F^{B\eta} &=& F^{B\eta_8}\cos\theta - F^{B\eta_0}\sin\theta\ ,\nn\\
F^{B\eta'} &=& F^{B\eta_8}\sin\theta + F^{B\eta_0}\cos\theta\ ,\ea

\noindent for $F^{B\pi}(0)=0.333$ and the mixing angle $\theta=-15.4^\circ$ \cite{feldmann},
we obtain the values $F^{B\eta}(0)=0.181$ and $F^{B\eta'}(0)=0.148$.

The $\rho^0-\omega$ mixing and isospin breaking effects are introduced in hadronic matrix
elements $B\to \rho^0$, following Ref. \cite{gabriel}. In the limit of isospin symmetry 
isospin eigenstates $\rho^I$ and $\omega^I$ expressed in the flavor basis are
$\rho^I=(u\bar u-d\bar d)/\sqrt{2}$ and $\omega^I=(u\bar u+d\bar d)/\sqrt{2}$. The physical
states $\rho^0$ and $\omega$ are expressed in term of $\rho^I$ and $\omega^I$ by

\ba &&|\rho^0\rangle = |\rho^I\rangle + \epsilon |\omega^I\rangle =
\frac{1}{\sqrt{2}}(1+\epsilon)|u\bar u\rangle +
\frac{1}{\sqrt{2}}(-1+\epsilon)|d\bar d\rangle\nn\\
&& |\omega \rangle = |\omega^I\rangle - \epsilon'|\rho^I\rangle =
\frac{1}{\sqrt{2}}(1-\epsilon')|u\bar u\rangle +
\frac{1}{\sqrt{2}}(1+\epsilon')|d\bar d\rangle\ ,\ea

\noindent where the numerical values for mixing parameters are $(1+\epsilon)=(0.092+0.016i)$ 
and $(1-\epsilon')=(1.011+0.030i)$. The hadronic matrix elements for the $B\to \rho^0$ and 
$B\to \omega$ transitions including isospin effects change by the factor $(1+\epsilon)$ and 
$(1-\epsilon')$, respectively. The effect in $B\to \omega$ transitions is negligible and it 
is not included in branching ratios predictions.

In the LCSR approach form factors for B decays are given in terms of the correlation
function of the weak current and the current with quantum numbers of B meson, evaluated
between the vacuum and a pseudoscalar or a vector meson. The like cone expansion allows to
calculate in the large virtualities of these currents. In the short virtualities regime,
the LCSR approach depends on the factorization of correlation function into 
nonpertubative and universal hadron function amplitudes which are convoluted with 
process depend amplitudes.

In Ref. \cite{ball} form factors for $B\to P$ and $B\to V$ transitions are calculated in the 
LCSR approach. In Table III form factors values at zero momentum transfer are showed, 
for the set $2$ of parameters, taken from Ref. \cite{ball}. For the $q^2$ dependency of the 
form factors we use the fit parametrization done in Ref. \cite{ball}, valid for the full 
kinematic regime.

\begin{table}[ht]
{\small TABLE III.~Form factors at zero momentum transfer for $B\to P$ and $B\to V$
transitions, evaluated in the WSB quark model \cite{wsb} and LCSR \cite{ball}.}
\begin{center}
\begin{tabular}{l c c c c c} \hline \hline
Transition  & $F_1=F_0$ & $V$ & $A_1$ & $A_2$ & $A_3=A_0$ \\
\hline
$B\to\pi$    & 0.333 [0.258]& & & &  \\
$B\to K$     & 0.379 [0.331]& & & & \\
$B\to\eta$   & 0.168 [0.275]& & & & \\
$B\to\eta'$  & 0.114 [-]& & & & \\
$B\to\rho$   & & 0.329 [0.323]& 0.283 [0.242]& 0.283 [0.221]& 0.281 [0.303]\\
$B\to\omega$ & & 0.232 [0.311]& 0.199 [0.233]& 0.199 [0.181]& 0.198 [0.363]\\
$B\to K^*$   & & 0.369 [0.293]& 0.328 [0.219]& 0.331 [0.198]& 0.321 [0.281]\\
\hline \hline
\end{tabular}
\end{center}
\end{table}

\subsection{Form factors for $B\to A$ in the ISGW2 model}

The ISGW2 model is based in a nonrelativistic constituent quark representation. In
the original ISGW model \cite{isgw89} form factors only depend on the maximum
momentum transfer, $q^2=q^2_m$. In this model form factor dependence is proportional to
$\exp[-(q^2_m-q^2)]$, consequently the form factors diminish exponentially as function of
$(q^2_m-q^2)$. This behavior has been improved in the ISGW2 model \cite{isgw95} by
expressing the $q^2$ dependence as a polynomial term which must be multiplied by a factor
which depends on the hyperfine mass. In addition, the improved model incorporates constrains
imposed by heavy quark symmetry, hyperfine distortions of wave functions and a more real
high recoil behavior.

We have made use of the ISGW2 model \cite{isgw95} to determine form factors for $B\to A$
transitions. The vector and axial part of matrix element for these transitions are
parametrized as

\ba \langle A(p_A,\epsilon)|(V_\mu-A_\mu)| B(p_B)\rangle &\equiv& l\,
\epsilon_\mu + (\epsilon \cdot p_B)\Big[c_+\,
(p_B + p_A)_\mu + c_-\, (p_B - p_A)_\mu \Big]\nn\\
&&-i\, q\, \epsilon_{\mu\nu\alpha\beta}\, \epsilon^\nu\,
(p_B+p_A)^\alpha (p_B-p_A)^\beta\ ,\ea

\noindent where $A(p_A,\epsilon)$ is a $^3P_1$ axial-vector meson.  For the $^1P_1$
axial-vector meson we change in the above matrix element $l$, $c_+$, $c_-$ and $q$ by
$r$, $s_+$, $s_-$ and $v$, respectively.

Considering $B\to A$ transitions, at quark level $b\to q_1$, axial-vector meson $A$
has the quark content $q_1 \bar q_2$, being $q_2$ the spectator quark. Thus, form factors
defined in the ISGW2 model have the following expressions

\ba l &=& -\tilde m_B\beta_B\left[\frac{1}{\mu_-}+
\frac{m_2\tilde m_A(\tilde\omega-1)}{\beta_B^2}
\left(\frac{5+\tilde\omega}{6 m_1}-\frac{1}{2\mu_-}\,\frac{m_2}{\tilde m_A}
\,\frac{\beta_B^2}{\beta^2_{BA}}\right)\right]F_5^{(l)}\nn\\
c_+ + c_- &=&  -\frac{m_2\tilde m_A}{2 m_1\tilde m_B\beta_B}
\left(1-\frac{m_1m_2}{2\tilde m_A\mu_-}\,\frac{\beta_B^2}{\beta^2_{BA}}\right)
F_5^{(c_+ + c_-)}\nn\\
c_+ - c_- &=& -\frac{m_2\tilde m_A}{2 m_1\tilde m_B\beta_B}
\left(\frac{\tilde\omega+2}{3}-\frac{m_1m_2}{2\tilde m_A\mu_-}\,
\frac{\beta_B^2}{\beta^2_{BA}}\right)F_5^{(c_+ - c_-)}\nn\\
q &=& -\frac{m_2}{2\tilde m_A \beta_B}\left(\frac{5+\tilde \omega}{6}\right)
F^{(q)}_5 \label{ff1}\ea

\noindent for the $^3P_1$ axial-vector meson and

\ba r &=& \frac{\tilde m_B\beta_B}{\sqrt{2}}\left[\frac{1}{\mu_+}
+\frac{m_2\tilde m_A}{3 m_1\beta^2_B}(\tilde\omega-1)^2\right]F_5^{(r)}\nn\\
s_+ + s_- &=& \frac{m_2}{\sqrt{2}\tilde m_B\beta_B}\left( 1-\frac{m_2}{m_1}
+\frac{m_2}{2\mu_+}\,\frac{\beta_B^2}{\beta^2_{BA}}\right)F_5^{(s_++s_-)}\nn\\
s_+ - s_- &=& \frac{m_2}{\sqrt{2}m_1\beta_B}\left(\frac{4-\tilde\omega}{3}-
\frac{m_1m_2}{2\tilde m_A\mu_+}\,
\frac{\beta_B^2}{\beta^2_{BA}}\right)F_5^{(s_+-s_-)}\nn\\
v &=& \Bigg[\frac{\tilde m_B \beta_B}{4\sqrt{2}m_b m_1\tilde m_A}+
\frac{(\tilde\omega-1)}{6\sqrt{2}}\frac{m_2}{\tilde m_A \beta_B}\Bigg]F^{(v)}_5
\label{ff2}\ea

\noindent for the $^1P_1$ axial-vector meson. The $F^{(i)}_5$ factors in the above
expressions are defined by

\ba F_5^{(l)} &=& F_5^{(r)} = F_5\left(\frac{\bar m_B}{\tilde m_B}\right)^{1/2}
\left(\frac{\bar m_A}{\tilde m_A}\right)^{1/2},\nn\\
F_5^{(q)} &=& F_5^{(v)} = F_5\left(\frac{\bar m_B}{\tilde m_B}\right)^{-1/2}
\left(\frac{\bar m_A}{\tilde m_A}\right)^{-1/2} ,\nn\\
F_5^{(c_+ + c_-)} &=& F_5^{(s_+ + s_-)} = F_5\left(\frac{\bar m_B}
{\tilde m_B}\right)^{-3/2}\left(\frac{\bar m_A}{\tilde m_A}\right)^{1/2},\nn\\
F_5^{(c_+ - c_-)} &=& F_5^{(s_+ - s_-)} = F_5\left(\frac{\bar m_B}
{\tilde m_B}\right)^{-1/2}\left(\frac{\bar m_A}{\tilde m_A}\right)^{-1/2}\ ,\ea

\noindent and the $F_n$ function by

\ba F_n &=& \left(\frac{\tilde m_A}{\tilde m_B}\right)^{1/2}
\left(\frac{\beta_B\beta_A}{\beta_{BA}}\right)^{n/2}
\left[1+\frac{1}{18}r^2(t_m-t)\right]^{-3},\ea

\noindent where

\ba r^2=\frac{3}{4 m_b m_1}+\frac{3m_2^2}{2\bar m_B\bar m_A\beta_{BA}^2}
+\frac{1}{\bar m_B\bar m_A}\left(\frac{16}{33-2n_f}\right)
\ln\left[\frac{\alpha_s(\mu_{\rm QM})}{\alpha_s(m_1)}\right]\ .\ea

The parameters $m_1$ and $m_2$ are masses of quarks $q_1$ and $q_2$, $\bar m$ is the
hyperfine averaged mass, $\tilde m$ is the sum of the masses of constituent quarks,
$t_m=(m_B-m_A)^2$ is the maximum momentum transferred, $n_f$ is the number of active
flavors at the $b$ scale and $\alpha_s(\mu)$ is the QCD coupling at the $\mu$ scale.
The parameters $\beta_B$, $\beta_A$ are obtained from the model, see Ref. \cite{isgw95}.
Moreover, we use the definitions

\ba \mu_{\pm} = \left(\frac{1}{m_1}\pm\frac{1}{m_b}\right)\ ,\
\tilde\omega = \frac{t_m-t}{2\bar m_B\bar m_A} + 1 \ea

\noindent and $\beta^2_{BA}=1/2(\beta^2_B+\beta^2_A)$.

In Table IV, we list values of form factors at momentum transferred $t=m^2_\pi$.
Form factors are functions of momentum transferred $t=(p_B-p_A)^2$, see Eqs.
(\ref{ff1}) and (\ref{ff2}). In general, form factors vary from $m^2_\pi$ to
$m^2_{K_1(1400)}$, in just only $4\%$. In addition, values for the form factors
depend strongly on the parameters $\beta_B=0.43$ and $\beta=0.28$ calculated in the model. 

We calculate form factors for $B\to K_{1A}$ and
$B\to K_{1B}$ transitions in $SU(3)$ base. Branching ratios are calculated with physical
form factors, which are obtained from the mixing, see Eq. (\ref{Kmatrix}).

\begin{table}[ht]
{\small TABLE IV.~Form factors at momentum transfer $t=m^2_\pi$ for $B\to A$ transitions,
evaluated in the ISGW2 model \cite{isgw95}.}
\begin{center}
\begin{tabular}{l c c c c} \hline \hline
Transition    & $q$ & $l$ & $c_+$ & $c_-$ \\
\hline
$B\to a_1$    & -0.0417 & -1.7469 & -0.0101 & -0.0012 \\
$B\to f_1$    & -0.0427 & -1.7603 & -0.0103 & -0.0012 \\
$B\to K_{1A}$ & -0.0593 & -1.8567 & -0.0155 & -0.0011 \\
              &   $v$     &   $r$  &   $s_+$ &  $s_-$  \\
$B\to b_1$    & 0.0319  & 0.9404  & 0.0177  & -0.0082 \\
$B\to h_1$    & 0.0324  & 0.9214  & 0.0134  & -0.0051 \\
$B\to K_{1B}$ & 0.0323  & 0.8956  & 0.0275  & -0.0124 \\
\hline \hline
\end{tabular}
\end{center}
\end{table}

To compare form factor values for $B\to A$ transitions with those of $B\to V$ transitions, 
we can define $B\to A$ transitions in same basis as the used in the BSW model, 
see Eq. (\ref{ffvector}). We change the symbols for the form factors $V$ and $A_{0,1,2}$
by  $A$ and $V_{0,1,2}$, respectively. These form factors are related to form factors
in the ISGW2 model by

\ba A(q'^2) &=& -(m_B+m_A)q(q'^2) ,\ \ \ V_1(q'^2) = \frac{l(q'^2)}{(m_B+m_A)},\ \ \  
V_2(q'^2) = -(m_B+m_A)c_+(q'^2)\nn\\
&& V_0(q'^2) = \frac{1}{2m_A}[l(q'^2)+(m^2_B-m^2_A)c_+(q'^2)+q'^2 c_-(q'^2)]\ .\ea

In Table V, we show form factor values for $B\to A$ transitions at momentum transferred
$t=m^2_\pi$, which correspond to those of Table IV.

\begin{table}[ht]
{\small TABLE V.~Form factors $V_{0,1,2,3}$ and $A$ at momentum transfer $t=m^2_\pi$ 
for $B\to A$ transitions, evaluated in the ISGW2 model \cite{isgw95}.}
\begin{center}
\begin{tabular}{l c c c c} \hline \hline
Transition    & $A$ & $V_1$ & $V_2$ & $V_3=V_0$ \\
\hline
$B\to a_1$    & 0.271  & -0.268 & 0.068  & -0.818 \\
$B\to f_1$    & 0.280  & -0.268 & 0.068  & -0.792 \\
$B\to K_{1A}$ & 0.389  & -0.283 & 0.102  & -0.890 \\        
$B\to b_1$    & -0.208 & 0.145  & -0.115 & 0.572 \\
$B\to h_1$    & -0.209 & 0.143  & -0.086 & 0.546 \\
$B\to K_{1B}$ & -0.216 & 0.134  & -0.184 & 0.573 \\
\hline \hline
\end{tabular}
\end{center}
\end{table}

\section{Amplitudes and branching ratios}

Let us present a comparison between $B\to V$ and $B\to A$ transitions, which seems
straightforward. First, we can see, from sections 2, 4, and 6 in  appendix B in Ref.
\cite{isgw89}, that parametrizations of $\langle V|J_\mu|B\rangle$ and
$\langle A|J_\mu|B\rangle$ are only different by a global sign, with the substitution of
the form factors $f\leftrightarrow l, r$,
$a_\pm \leftrightarrow c_\pm, s_\pm$, $g \leftrightarrow q, v$. This is because behavior
of currents $V_\mu$ and $A_\mu$ are interchanged. Moreover, this implies that expressions
for decay amplitudes and decay rates, at tree level, for the processes $B\to VP$, $VV$ and
$B\to AP$, $AV$, and $AA$, respectively, are identical.

On the other hand, the situation is different when tree and penguin contributions are
considered. The expressions of decay amplitudes for processes $B\to AP$ (see appendix A)
and $B\to VP$ (see appendices in Refs. \cite{ali, chen99}) are equal when only QCD parameters
$a_3$, $a_4$, $a_9$ and $a_{10}$ contribute. When QCD parameters $a_6$ and $a_8$ contribute
then the linear combination ($z a_6 + y a_8$) is affected by a global sign and $1/(m_b-m_q)$,
which is a factor of this linear combination, changes by $1/(m_b+m_q)$. Relevant
contributions in penguin sector are coefficients $a_4$ and $a_6$ (see appendix A); in $B \to A$
transitions $a_6$ and $a_4$ have same sign. This fact implies that these terms are summed
so their contribution increase. In $B \to V$ transitions, these terms have different sign thus
their contribution decrease. The contributions corresponding to $a_5$ and $a_7$ change sign
when the axial-vector or the vector meson arises from vacuum, but they are not affected if the
pseudoscalar meson is produced from vacuum.

Now we are going to compare penguin contributions to decay amplitudes ${\cal M}(B\to AV)$ 
(see appendix B) with the ones ${\cal M}(B\to VV)$ (showed in appendices F and G in Ref. 
\cite{chen99}): (i) in both cases contribution of parameters $a_6$ and $a_8$ does not appear; 
(ii) sign of contribution given by parameters $a_5$ and $a_7$ changes when one goes from 
$B\to V,A$ to $B\to V,V$; (iii) in modes $B\to V,V^{charged}$ and $B\to V,A^{charged}$ always
appears the contribution $(a_4+a_{10})$. 

In Table VI, we have summarized penguin contributions to decay amplitudes for modes $B\to AP$
displayed in appendix A, without including $P=\eta^{(')}$. These decay amplitudes can be
classified in two groups from these contributions. The first group is integrated by decays
where a charged meson in final state is produced from vacuum and penguin contribution is given
by the linear combination $(a_4+a_{10})+\alpha(a_6+a_8)R$, i.e., parameters $a_{even}$ only
contribute to this group. Additionally, in this group we find two cases with $\alpha=0$ and
$\alpha=1$, which correspond to modes $B\to P,A^{charged}$ and $B\to A,P^{charged}$,
respectively. Here the notation $B\to M_1,M_2$ means that meson $M_2$ can be factorized out
under the factorization approximation.

The second group is integrated by decays where a neutral meson is factorized out under
factorization approximation independently if it is pseudoscalar or axial-vector. Penguin
contribution is given by the linear combination
$\alpha_1(a_4 -a_{10}/2)+\alpha_2(a_6-a_8/2)R+\alpha_3(a_7-a_9)+\alpha_4(a_3-a_5)$. Pure
penguin contributions belong to this group and have contributions of $a_{even}$. They arise
when the axial-vector meson or the pseudoscalar meson is a neutral strange meson and, of
course, it is produced from vacuum. QCD parameters $a_4$, $a_6$, $a_8$ and $a_{10}$ contribute
when the pseudoscalar $K^0$ meson is factorized out under factorization approximation, $a_4$
and $a_{10}$ when the axial-vector $K_1^0$ meson arises from vacuum. Note, that in general,
decays $B\to P,A$ do not have contributions from $a_6$ and $a_8$.

In Table VII, we have classified penguin contributions to decay amplitudes ${\cal M}(B\to AV)$
which are showed in appendix B. There are two types: in one of them the linear combination
$\alpha_1(a_4 +a_{10})+\alpha_2(a_7\pm a_9)$ contribute. It occurs when decays
$B\to A, V^{charged}$ or $B\to V,A^{charged}$ are produced, i.e., when a charged meson in the
final state arises from vacuum; in the other case, a neutral meson is factorized out under
factorization approximation and the linear combination
$\beta_1(a_4-a_{10}/2)+\beta_2(a_3\pm a_5)+\beta_3(a_7\pm a_9)$ contributes. Pure penguin
contributions belong to it. Parameters $a_{odd}$ contribute to decay amplitude of pure penguin
modes $\bar B^0\to a^0_1f_1$ and $\bar B^0\to a^0_1\phi$. Like decays $B\to P,A$, in general, 
decays $B\to AV$ do not have contributions from $a_6$ and $a_8$.

Penguin contributions of decay amplitudes ${\cal M}(B\to AA)$ (see appendix C) can be
classified in a similar way. There are two groups. In one of them a charged meson is factorized
out under factorization scheme and only $a_4$ and $a_{10}$ parameters contribute by means of
the linear combination $(a_4 +a_{10})$. In the other group a neutral meson is produced from
vacuum. In this case the linear combination
$\zeta_1(a_4-a_{10}/2)+\zeta_2(a_3-a_5)+\zeta_3(a_7-a_9)$ contribute. In Table VIII, we display
the respective coefficients $\zeta_i$. Again, pure penguin decays are in this group.

\begin{table}[ht]
{\small TABLE VI. Coefficients of the linear combinations
$(a_4+a_{10})+\alpha(a_6+a_8)R$  and $\alpha_1(a_4 -a_{10}/2)+
\alpha_2(a_6-a_8/2)R+\alpha_3(a_7-a_9)+\alpha_4(a_3-a_5)$ corresponding  to penguin
contribution of decay amplitudes ${\cal M}(B\to AP)$ without $P=\eta^{(`)}$. The
coefficient R is given by $R=2m_P^2/(m_1+m_2)(m_b-m_3)$}.
\begin{center}
\begin{tabular}{l c c c c c} \hline \hline
Decays  & $\alpha$ & $\alpha_1$ & $\alpha_2$ & $\alpha_3$& $\alpha_4$ \\
\hline
$\bar B^0 \to \pi^+,a_1^- $; $B^-\to \pi^0,a_1^- $;
$\bar B^0 \to \pi^+,K_1^- $; $B^-\to \pi^0,K_1^- $
& 0 &  & & & \\
$\bar B^0 \to a_1^+,\pi^-$; $B^-\to a_1^0,\pi^-$; $\bar B^0 \to a_1^+,K^-$;
$B^-\to a_1^0,K^-$; $B^-\to f_1,\pi^-$; $B^-\to f_1,K^-$
& 1& & & & \\ \hline
$\bar B^0 \to \pi^0,f_1$; $B^- \to \pi^-,f_1$
& & 1 & 0 & -1/2 & 2 \\
$\bar B^0 \to a_1^0,\pi^0$; $B^- \to a_1^-,\pi^0$;
$\bar B^0 \to f_1,\pi^0$ & & $\pm1$ & $\pm1$ & $\pm3/2$ & 0 \\
$\bar B^0 \to \pi^0,a_1^0$; $B^- \to \pi^-,a_1^0$ & & -1 & 0 & -3/2 & 0 \\
$\bar B^0 \to \bar K^0,f_1$; $B^- \to K^-,f_1$ & & 0 & 0 & -1/2 & 2 \\
$\bar B^0 \to f_1,\bar K^0$; $B^- \to a_1^-, \bar K^0$; $\bar B^0
\to a_1^0,\bar K^0$; $\bar B^0 \to \bar K_1^0,K^0$;
$B^- \to K_1^-,K^0$ & & 1 & 1 & 0 & 0 \\
$\bar B^0 \to \bar K^0,a_1^0$; $B^- \to K^-,a_1^0$; $B^- \to K_1^-,\pi^0$;
$\bar B^0 \to \bar K_1^0,\pi^0$ & & 0 & 0 & -3/2 & 0 \\
$\bar B^0 \to \pi^0,K_1^0$; $B^- \to \pi^-, \bar K_1^0$; $\bar B^0
\to \bar K^0,K_1^0$;
$B^- \to K^-,K_1^0$ & & 1 & 0 & 0 & 0 \\
\hline \hline
\end{tabular}
\end{center}
\end{table}

\begin{table}[ht]
{\small TABLE VII. Coefficients of the linear combinations
$\alpha_1(a_4 +a_{10})+\alpha_2(a_7\pm a_9)$ and
$\beta_1(a_4-a_{10}/2)+\beta_2(a_3\pm a_5)+\beta_3(a_7\pm a_9)$ corresponding to penguin
contribution of decay amplitudes ${\cal M}(B\to AV)$.}
\begin{center}
\begin{tabular}{l c c c c c } \hline \hline
Decays  & $\alpha_1$ & $\alpha_2$ &  $\beta_1$& $\beta_2$& $\beta_3$ \\
\hline
$\bar B^0\to \rho^+,a_1^- (K_1^-)$; $\bar B^0\to a_1^+,\rho^-$;
$\bar B^0(B^-)\to a_1^+(a_1^0),K^{*-} $;\\ $B^-\to \omega,a_1^-(K_1^-)$;
$B^-\to f_1,\rho^-(K^{*-}) $; $B^- \to \rho^0, K_1^-$   &1& 0 & & & \\
\hline $B^-\to a_1^-,\rho^0 $; $B^-\to \rho^-,a_1^0(\bar K_1^0) $;
$\bar B^0\to a_1^0,\bar K^{*0}$; $B^-\to a_1^-,\bar K^{*0}$; $\bar
B^0\to f_1(\bar K_1^0),\bar K^{*0}$; $\bar B^0 \to \omega, \bar
K_1^0$
 & & &$\pm 1$ &0  &0  \\
$\bar B^0\to a_1^0,\rho^0 $; $\bar B^0\to \rho^0,a_1^0 $; $\bar
B^0\to \omega,a_1^0 $;
$\bar B^0\to f_1,\rho^0 $ &&  &-1&0 &$\pm 3/2$  \\
$\bar B^0\to \bar K^{*0},a_1^0 $; $B^-\to  K^{*-},a_1^0 $; $\bar B^0
\to \bar K_1^0, \rho^0$; $B^- \to K_1^-, \rho^0$
 &&  &0&0 &$- 3/2$ \\
$\bar B^0\to a_1^0,\omega $; $B^-\to a_1^-,\omega $; $\bar B^0\to
\rho^0,f_1 $; $B^-\to \rho^-,f_1$; $\bar B^0\to f_1,\omega $; $\bar
B^0\to \omega,f_1 $& & & 1 & 2&
$\pm 1/2$\\
$\bar B^0\to a_1^0, \phi $; $\bar B^0\to f_1,\phi $ & & & 0 & 1& $-1/2$\\
$\bar B^0\to K^{*0},f_1 $; $B^-\to K^{*-},f_1$;
$\bar B^0 \to \bar K_1^0, \omega$; $B^- \to K_1^-, \omega$ && &0&2 & $- 1/2$ \\
$\bar B^0\to \bar K_1^0,\phi $; $B^-\to K_1^-,\phi$ && &1&1 & $- 1/2$ \\
\hline \hline
\end{tabular}
\end{center}
\end{table}

\begin{table}[ht]
{\small TABLE VIII. Coefficients of the linear combination
$\zeta_1(a_4 -a_{10}/2)+\zeta_2(a_3 - a_5)+\zeta_3(a_7 - a_9)$ corresponding to penguin 
contribution of decay amplitudes ${\cal M}(B\to A, A^{neutral})$.}
\begin{center}
\begin{tabular}{l c c c} \hline \hline
Decays  & $\zeta_1$& $\zeta_2$& $\zeta_3$ \\
\hline
$\bar B^0\to a_1^0,a_1^0$; $\bar B^0\to f_1,a_1^0$   & -1 & 0 & -3/2 \\
$\bar B^0\to a_1^0,f_1$; $B^- \to a_1^-,f_1$ & 1 & 2 & -1/2 \\
$\bar B^0\to a_1^0,\bar K_1^0$; $B^- \to a_1^-,\bar K_1^0$;
$\bar B^0\to f_1,\bar K_1^0$ & 1 & 0 & 0 \\
$\bar B^0\to \bar K_1^0,a_1^0$; $B^- \to K_1^-,a_1^0$ & 0 & 0 & -3/2 \\
$\bar B^0\to \bar K_1^0,f_1$; $B^- \to K_1^-,f_1$ & 0 & 2 & -1/2 \\
\hline \hline
\end{tabular}
\end{center}
\end{table}

In appendices we give explicitly the amplitudes to the processes studied in terms of
form factors for $B\to P$, $B\to V$ and $B\to A$ transitions. In appendix A we have 
a common factor $(\epsilon^* \cdot p_B)$ which is not included in the expressions to 
simplify. And we use the symbol $K_1$ to indicate the axial-vector mesons $K_1(1270)$
or $K_1(1400)$. In appendices we have the factor $G_F/\sqrt{2}$ common to all
amplitudes.

It is straightforward to calculate the branching ratios from amplitudes and input parameters.
However, here we give general expressions which are useful in decay rates estimations. The
decay rate formula for $B\to AX$ decays is, in general, given by

\ba \Gamma(B\to AX) &=& \frac{p_c}{8\pi m^2_B}|{\cal M}(B\to AX)|^2 \ea

\noindent where $p_c=\lambda^{1/2}(m^2_B,m^2_A,m^2_V)/2m_B$ is the momentum of decay particle
in the rest frame of $B$ meson and $X$ can be $P$, $V$ or $A$ and 
$\lambda(a,b,c)\equiv a^2+b^2+c^2-2(ab+ac+bc)$

For branching ratios of $B\to AP$ decays, we note that amplitude ${\cal M}(B\to AP)$ is 
proportional to $(\epsilon^*_A\cdot p_B)$. Thus amplitude squared is proportional to 
$|(\varepsilon^*_A\cdot p_B)|^2$, which is easily calculated. The general decay rate 
formula for $B\to AV$ decays is more involved, because the amplitude ${\cal M}(B\to AV)$ 
includes an interfering term. Since the long expressions we do not write here.

\section{Numerical results}

In this section we present our numerical results. In Tables IX-X, XI-XII, and XIII-XIV
we display branching ratios of $B\to AP$, $B \to AV$ and $B \to AA$ decays, respectively,
using the improved version of the ISGW quark model \cite{isgw95} for calculating form
factors for $B\to A$ transitions.

Branching ratios for $B\to AP$ decays, where $A$ is a $^3P_1$ nonstrange axial-vector
meson (see Table IX) are bigger than ones where $A$ is a $^1P_1$ nonstrange axial-vector
meson. The ratio $Br(B\to A(^3P_1)P)/Br(B\to A(^1P_1)P)$, where mesons $A(^3P_1)$ and
$A(^1P_1)$ have the same quark content, is $\sim 1.6-4.5$, except for a small number of
them. The mode $B^-\to a_1^-\bar K^0$, which is a pure penguin channel, is the most dominant
(its branching ratio of $84.1\times 10^{-6}$ is the biggest). Penguin contribution to this
mode is given by $a_{even}$ parameters. Other dominant decays are $\bar B^0\to a_1^+\pi^-$ and
$\bar B^0\to a_1^+K^-$, whose branching ratios are $74.3\times 10^{-6}$ and $72.2\times 10^{-6}$,
respectively. In these decays there is a destructive interference between penguin and
$W$-external or $W$-internal contributions. On the other hand, a similar situation is found
changing a $^3P_1$ meson by a $^1P_1$ meson with the same quark content (see fourth column in
Table IX). At experimental level there is not enough information. Our predictions for
$Br(\bar B^0\to a_1^-\pi^+)=36.7\times 10^{-6}$ and 
$Br(\bar B^0\to a_1^+\pi^-)=74.3\times 10^{-6}$ are consistent with the experimental average 
value $Br(\bar B^0\to a_1^{\mp}\pi^{\pm})=(40.9\pm 7.6)\times 10^{-6}$ \cite{nardulli06}.
This average includes BaBar and Belle results \cite{exp}. Finally, we want to mention that our
predictions are at the same order that the ones obtained by Laporta-Nardulli-Pham (see Tables V
and VI in Ref. \cite{nardulli06}), although our values are in general bigger, except in a few
modes.

In Table X, we show branching ratios for $B\to K_1P$ decays for two values
($\theta=32^\circ,58^\circ$) of the mixing angle $K_A-K_B$. The strange axial-vector meson
is $K_1(1270)$ or $K_1(1400)$. In this case, the most dominant decays are 
$B^-\to K_1^-(1400)\etap$, with branching ratios of ${\cal O}(10^{-5})$. On the other hand, 
branching ratios of modes
$\bar B^0\to K_1^-\pi^+$, $\bar B^0\to \bar K_1^0\pi^0$, $B^-\to K_1^-\pi^0$, 
$B^-\to \bar K_1^0\pi^-$, $\bar B^0\to K_1^0\bar K^0$ and $B^-\to K_1^0K^-$, where the $K_1$ 
meson can be $K_1(1270)$ or $K_1(1400)$, are not sensitive to the value of the mixing angle. 
On the contrary, branching ratios of $\bar B^0\to \bar K_1^0(1270)\etap(K^0)$ and 
$B^-\to K_1^-(1270)\etap(K^0)$ strongly depend on the mixing angle. The same decays but 
changing $K_1(1270)$ by $K_1(1400)$ are not very sensitive to this angle. Branching ratios of 
$B\to K_1(1270)P$ and $B\to K_1(1400)P$ are smaller with $\theta=32^\circ$ and 
$\theta=58^\circ$, respectively. Laporta-Nardulli-Pham, in Table IV, Ref. \cite{nardulli06}, 
displayed some of the branching ratios that we present in Table X. In general, both 
predictions agree.

In Table XI, we display branching ratios for $B\to AV$ decays with $A$ being a $^3P_1$ or a
$^1P_1$ nonstrange axial-vector meson. The most of these decays are suppressed. In general,
$Br(B\to A(^3P_1)V)$ is bigger than $Br(B \to A(^1P_1)V)$, where mesons $A(^3P_1)$ and
$A(^1P_1)$ have the same quark content. In this case, dominant decays are $B^-\to f_1\rho^-$,
$\bar B^0\to a_1^{\pm}\rho^{\mp}$. Their branching ratios are ${\cal O}(10^{-6})$. If we compare
Tables IX and XI we found that $Br(B\to A P(q_1\bar q_2))>Br(B\to A V(q_1\bar q_2))$.

In Table XII, we present branching ratios for $B\to K_1V$ decays for two values
($\theta=32^\circ,58^\circ$) of the mixing angle $K_A-K_B$. The strange axial-vector meson is
$K_1(1270)$ or $K_1(1400)$. These decays are, in general, suppressed. The dominant decays are
$\bar B^0\to \bar K_1^0(1400)K^{*0}$, $B^-\to K_1^-(1400)K^{*0}$ and $B^-\to K_1^0(1270)K^{*-}$.
Their branching ratios are ${\cal O}(10^{-6})$. On the other hand, the branching ratios of
modes $\bar B^0\to K_1^-\rho^+$, $B^-\to\bar K_1^0\rho^-$, $B^-\to K_1^-\omega$,
$\bar B^0\to K_1^0\bar K^{*0}$ and $B^-\to K_1^0K^{*-}$, with $K_1=K_1(1270),K_1(1400)$, are
not sensitive to the value of the mixing angle. On the contrary, branching ratio of
$\bar B^0\to \bar K_1^0(1270)K^{*0}$ and $B^-\to K_1^-(1270)K^{*0}$ strongly depend on the value
of $\theta$. The predictions obtained in Ref. \cite{chen05} (see Tables II and III) for
$B\to K_1(1270)\phi$ and $B\to K_1(1400)\phi$ and our predictions for these modes do not agree,
except for the case with $N_c^{eff}=\infty$ and $\theta=58^\circ$ (with $\mu=2.5$ GeV or
$\mu=4.4$ GeV). In this case the respective branching ratios are ${\cal O}(10^{-7})$.

The branching ratios for $B\to b_1(h_1)P$ and $B\to b_1(h_1)V$ decays are calculated in the 
$SU(3)$ symmetry limit, i.e. $f_{b_1}=f_{h_1}=0$. In Table XIII we include the modes which 
change values with respect to Tables IX and XI. The branching ratios for 
$\bar B^0\to b^-_1\pi^+$ and $\bar B^0\to b^-_1\rho^+$ decays are zero since the amplitude 
is proportional to the decay constant $f_{b_1}$. The branching ratios for 
$B^-\to b^-_1\pi^0$, $B^-\to b^-_1\etap$, $B^-\to b^-_1\omega$ decays decrease by one 
order of magnitude and for $B^-\to h_1K^{*-}$, $\bar B^0\to h_1K^{*0}$ decays decrease by 
one half order of magnitude with respect to the values in Tables IX and XI. 
The modes $B^-\to b^-_1\eta'$, $B^-\to b^0_1\omega$ and 
$\bar B^0\to h_1\rho^0$ increase by one order of magnitude. 

Moreover, in Tables IX-XII, between brackets we show values corresponding to the branching
ratios $B\to AP$ and $B\to AV$, where $B\to P(V)$ transitions are calculated in the LCSR 
approach. The values with the symbol $[-]$ represent branching ratios which basically have 
equal value with respect to the calculated with the WSB model. In general, the branching 
ratios are smaller compared with the calculated with the BSW model. The branching ratios for 
$B^-\to a^-_1\omega$,  $B^-\to b^-_1\omega$ and $B\to K_1\omega$ decays increase their 
values, because in the LCSR approach the form factors for $B\to \omega$ transitions are 
bigger compared with the BSW model.

In Table XIV, we present branching ratios for five $B\to AA$ decays, where $A$ is a
nonstrange axial-vector meson. The branching ratio of $\bar B^0\to a_1^-a_1^+$ is
${\cal O}(10^{-6})$. In this group, this decay is dominant. From Tables XI (see second column)
and XIV we conclude that $Br(B\to a_1^-V(q_1\bar q_2)\sim Br(B\to a_1^-A(q_1\bar q_2))$,
where $V$ and $A$ are nonstrange mesons.

In Table XV, we show branching ratios for $B\to K_1A$ decays for two values
($\theta=32^\circ,58^\circ$) of the mixing angle $K_A-K_B$. The strange axial-vector meson is
$K_1(1270)$ or $K_1(1400)$. Branching ratios of the decays $\bar B^0\to K_1^-a_1^+$ and
$B^-\to \bar K_1^0 a_1^-$ are not sensitive to the mixing angle. In this group, 
$Br(\bar B^0\to K_1^-(1270)a_1^+)\sim 10^{-7}$ is the biggest. From Tables XII and XV we 
conclude that $Br(B\to K_1\rho)\sim Br(B\to K_1a_1)$.

Finally, in Table XVI, we present a summary about experimental information given in Ref.
\cite{pdg2006} for branching ratios of some charmless $B \to AP, AV, AA$ decays. In general,
bounds for these branching ratios are $<(10^{-3}-10^{-4})$. There is a similar situation
for charmed and charmonium $B$ decays \cite{pdg2006}.

\begin{table}[ht]
{\small TABLE IX. Branching ratios (in units of $10^{-6}$) of $B\to A P$ decays,
where $A$ is a nonstrange axial-vector meson, using the ISGW2 form factors for 
$B\to A$ transitions and BSW [LCSR] for $B\to P$ transitions.}
\begin{center}
\begin{tabular}{l c l c}\hline \hline
Mode\, \, \, \,  & ${\cal B}$  & Mode\, \, \, \,  & ${\cal B}$  \\
\hline
$\bar B^0\to a^-_1 \pi^+$ & 36.7 [23.5] & $\bar B^0\to b^-_1 \pi^+$ & 4.4 [2.8] \\
$\bar B^0\to a^+_1 \pi^-$ & 74.3 [-] & $\bar B^0\to b^+_1 \pi^-$ & 36.2 [-] \\
$\bar B^0\to a^0_1 \pi^0$ & 0.27 [-] &  $\bar B^0\to b^0_1 \pi^0$ & 0.15 [-]  \\
$B^-\to a^-_1 \pi^0$      & 13.6 [7.8] &  $B^-\to b^-_1 \pi^0$      & 4.2 [3.1] \\
$B^-\to a^0_1 \pi^-$      & 43.2 [-] &  $B^-\to b^0_1 \pi^-$      & 18.6 [-] \\

$\bar B^0\to a_1^0 \eta$  & 0.54 [-] & $\bar B^0\to b_1^0 \eta$  & 0.17 [-] \\
$B^-\to a_1^- \eta$       & 13.4 [9.1] & $B^-\to b_1^- \eta$       & 0.88 [0.51] \\

$\bar B^0\to a_1^0 \eta'$ & 0.09 [-] &  $\bar B^0\to b_1^0 \eta'$ & 0.02 [-] \\
$B^-\to a_1^- \eta'$      & 13.6 [10.1] &  $B^-\to b_1^- \eta'$      & 0.02 [0.001] \\

$\bar B^0\to a^0_1 \bar K^0$ & 42.3 [-]&  $\bar B^0\to b^0_1 \bar K^0$ & 19.3 [-] \\
$\bar B^0\to a^+_1 K^-$      & 72.2 [-]&  $\bar B^0\to b^+_1 K^-$      & 35.7 [-] \\
$ B^-\to a^0_1 K^-$      & 43.4 [-]&  $ B^-\to b^0_1 K^-$      & 18.1 [-] \\
$ B^-\to a^-_1 \bar K^0$ & 84.1 [-]&  $ B^-\to b^-_1 \bar K^0$ & 41.5 [-] \\

$\bar B^0\to f_1 \pi^0$ & 0.47 [-]&  $\bar B^0\to h_1 \pi^0$ & 0.16 [-] \\
$B^-\to f_1 \pi^-$      & 34.1 [-]&  $B^-\to h_1 \pi^-$      & 18.6 [-] \\

$\bar B^0\to f_1 \eta$  & 37.1 [-] &  $\bar B^0\to h_1 \eta$  & 18.2 [-] \\
$\bar B^0\to f_1 \eta'$ & 22.1 [-] &  $\bar B^0\to h_1 \eta'$ & 11.2 [-] \\

$\bar B^0\to f_1 \bar K^0$ & 34.7 [-]& $\bar B^0\to h_1 \bar K^0$ & 19.0 [-]\\
$B^-\to f_1 K^-$           & 31.1 [-]& $B^-\to h_1 K^-$       & 19.0 [-]\\
\hline \hline
\end{tabular}
\end{center}
\end{table}

\begin{table}[ht]
{\small TABLE X.  Branching ratios (in units of $10^{-6}$) of $B\to AP$ decays,
where $A$ is a strange axial-vector meson $K_1(1270)$ or $K_1(1400)$, using the
ISGW2 form factors for $B\to A$ transitions and BSW [LCSR] for $B\to P$ transitions.}
\begin{center}
\begin{tabular}{l c c l c c} \hline \hline
Mode\, \, \, \, & ${\cal B}$ $(32^\circ)$ & ${\cal B}$
$(58^\circ)$
& Mode\, \, \, \,  & ${\cal B}$ $(32^\circ)$  & ${\cal B}$$(58^\circ)$\\
\hline
$\bar B^0\to K^-_1(1270) \pi^+$      & 4.3 [2.8] & 4.3 [2.8] &
$\bar B^0\to K^-_1(1400) \pi^+$      & 2.3 [1.5] & 2.3 [1.5] \\
$\bar B^0\to \bar K^0_1(1270) \pi^0$ & 2.3 [1.5] & 2.1 [1.4] &
$\bar B^0\to \bar K^0_1(1400) \pi^0$ & 1.7 [1.3] & 1.6 [1.3] \\
$B^-\to K^-_1(1270) \pi^0$           & 2.5 [1.6] & 1.6 [0.9] &
$B^-\to K^-_1(1400) \pi^0$           & 0.67 [0.51] & 0.64 [0.55] \\
$B^-\to \bar K^0_1(1270) \pi^-$      & 4.7 [3.0] & 4.7 [3.0] &
$B^-\to \bar K^0_1(1400) \pi^-$      & 2.5 [1.7] & 2.5 [1.7] \\

$\bar B^0\to \bar K_1^0(1270) \eta$  & 1.5 [1.1] & 10.2 [9.8] &
$\bar B^0\to \bar K_1^0(1400) \eta$  & 52.8 [52.5] & 46.8 [46.6] \\
$B^-\to K_1^-(1270) \eta$            & 0.95 [0.65] & 20.7 [19.4] &
$B^-\to K_1^-(1400) \eta$            & 95.1 [93.3] & 84.8 [83.1] \\

$\bar B^0\to \bar K_1^0(1270) \eta'$ & 1.1 [0.8] & 9.4 [9.1] &
$\bar B^0\to \bar K_1^0(1400) \eta'$ & 51.4 [51.2] & 46.0 [45.8] \\
$B^-\to K_1^-(1270) \eta'$           & 0.53 [0.4] & 16.6 [15.6] &
$B^-\to K_1^-(1400) \eta'$           & 80.0 [78.5] & 71.9 [70.5] \\

$\bar B^0\to K^0_1(1270) \bar K^0$ & 0.20 [0.17] & 0.20 [0.17] &
$\bar B^0\to K^0_1(1400) \bar K^0$ & 0.11 [0.09] & 0.11 [0.09] \\
$\bar B^0\to \bar K^0_1(1270) K^0$ & 0.02 [-] & 0.70 [-] &
$\bar B^0\to \bar K^0_1(1400) K^0$ & 4.1 [-] & 3.6 [-] \\
$B^-\to K^-_1(1270) K^0$      & 0.02 [-] & 0.75 [-] &
$B^-\to K^-_1(1400) K^0$      & 4.4  [-] & 3.9  [-] \\
$B^-\to K^0_1(1270) K^-$      & 0.22 [0.18] & 0.22 [0.18] &
$B^-\to K^0_1(1400) K^-$      & 0.12 [0.10] & 0.12 [0.10] \\
\hline \hline
\end{tabular}
\end{center}
\end{table}

\begin{table}[ht]
{\small TABLE XI.  Branching ratios (in units of $10^{-6}$) of $B\to AV$ decays,
where $A$ is a nonstrange axial-vector meson, using the ISGW2 form factors
$B\to A$ transitions and BSW [LCSR] for $B\to V$ transitions.}
\begin{center}
\begin{tabular}{l c l c}\hline \hline
Mode\, \, \, \,  & ${\cal B}$  & Mode\, \, \, \,  & ${\cal B}$  \\
\hline
$\bar B^0\to a^-_1 \rho^+$ & 4.7 [3.5] & $\bar B^0\to b^-_1 \rho^+$ & 0.55 [0.41]\\
$\bar B^0\to a^+_1 \rho^-$ & 4.3 [-] & $\bar B^0\to b^+_1 \rho^-$ & 1.6 [-]\\
$\bar B^0\to a^0_1 \rho^0$ & 0.01 [0.009] & $\bar B^0\to b^0_1 \rho^0$ & 0.002 [-]\\
$B^-\to a^-_1 \rho^0$ & 3.0 [2.3]& $B^-\to b^-_1 \rho^0$ & 0.36 [0.27]\\
$B^-\to a^0_1 \rho^-$ & 2.4 [-]& $B^-\to b^0_1 \rho^-$ & 0.86 [-]\\

$\bar B^0\to a_1^0 \omega$ & 0.003 [0.02]& $\bar B^0\to b_1^0 \omega$ & 0.004 [0.001]\\
$B^-\to a_1^- \omega$      & 2.2 [5.1]& $B^-\to b_1^- \omega$      & 0.38 [0.46]\\

$\bar B^0\to a_1^0 \phi$ & 0.0005 [-] & $\bar B^0\to b_1^0 \phi$ & 0.0002 [-] \\
$B^-\to a_1^- \phi$      & 0.001 [-] & $B^-\to b_1^- \phi$      & 0.0004 [-] \\

$\bar B^0\to a^0_1 K^{*0}$ & 0.64 [0.61] & $\bar B^0\to b^0_1 K^{*0}$ & 0.15 [-] \\
$\bar B^0\to a^+_1 K^{*-}$ & 0.92 [-] & $\bar B^0\to b^+_1 K^{*-}$ & 0.32 [-] \\
$\bar B^-\to a^0_1 K^{*-}$ & 0.86 [0.81] & $\bar B^-\to b^0_1 K^{*-}$ & 0.12 [0.13] \\
$\bar B^-\to a^-_1 \bar K^{*0}$ & 0.51 [-]& $\bar B^-\to b^-_1 \bar K^{*0}$ & 0.18 [-] \\

$\bar B^0\to f_1 \rho^0$ & 0.03 [-] & $\bar B^0\to h_1 \rho^0$ & 0.002 [-] \\
$B^-\to f_1 \rho^-$      & 4.9  [-] & $B^-\to h_1 \rho^-$      & 1.6 [-] \\

$\bar B^0\to f_1 \omega$ & 0.02 [-] & $\bar B^0\to h_1 \omega$ & 0.005 [-] \\
$\bar B^0\to f_1 \phi$   & 0.0005 [-] & $\bar B^0\to h_1 \phi$   & 0.0002 [-] \\

$\bar B^0\to f_1 K^{*0}$ & 0.43 [0.42]& $\bar B^0\to h_1 K^{*0}$ & 0.35 [0.32] \\
$B^-\to f_1 K^{*-}$      & 0.45 [0.48]& $B^-\to h_1 K^{*-}$      & 0.50 [0.48] \\
\hline \hline
\end{tabular}
\end{center}
\end{table}

\begin{table}[ht]
{\small TABLE XII. Branching ratios (in units of $10^{-6}$) of $B\to AV$ decays,
where $A$ is a strange axial-vector meson $K_1(1270)$ or $K_1(1400)$, using the ISGW2
form factors for $B\to A$ transitions and BSW [LCSR] for $B\to V$ transitions.}
\begin{center}
\begin{tabular}{l c c l c c}\hline \hline
Mode\, \, \, \, & ${\cal B}$ $(32^\circ)$ & ${\cal B}$ $(58^\circ)$
& Mode\, \, \, \,  & ${\cal B}$ $(32^\circ)$  & ${\cal B}$$(58^\circ)$\\
\hline
$\bar B^0\to K^-_1(1270) \rho^+$      & 0.62 [0.45] & 0.62 [0.45] &
$\bar B^0\to K^-_1(1400) \rho^+$      & 0.45 [0.31] & 0.45 [0.31] \\
$\bar B^0\to \bar K^0_1(1270) \rho^0$ & 0.001 [-] & 0.02 [-] &
$\bar B^0\to \bar K^0_1(1400) \rho^0$ & 0.05 [-] & 0.04 [-] \\
$B^-\to K^-_1(1270) \rho^0$           & 0.10 [0.03] & 0.05 [0.07] &
$B^-\to K^-_1(1400) \rho^0$           & 0.02 [0.01] &  0.02 [0.01] \\
$B^-\to \bar K^0_1(1270) \rho^-$      & 0.001 [-] & 0.001 [-] &
$B^-\to \bar K^0_1(1400) \rho^-$      & 0.001 [0.0006] & 0.001 [0.0006] \\

$\bar B^0\to \bar K_1^0(1270) \omega$ & 0.0002 [0.001] & 0.001 [0.003] &
$\bar B^0\to \bar K_1^0(1400) \omega$ & 0.004 [0.005] & 0.003 [0.007] \\
$B^-\to K_1^-(1270) \omega$           & 0.06 [0.16] & 0.07 [0.15] &
$B^-\to K_1^-(1400) \omega$           & 0.06 [0.07] & 0.06 [0.07] \\

$\bar B^0\to \bar K_1^0(1270) \phi$ & 0.004 [-] & 0.25 [-] &
$\bar B^0\to \bar K_1^0(1400) \phi$ & 0.87 [-] & 0.66 [-] \\
$B^-\to K_1^-(1270) \phi$           & 0.004 [-] & 0.27 [-] &
$B^-\to K_1^-(1400) \phi$           & 0.93 [-] & 0.69 [-] \\

$\bar B^0\to K^0_1(1270) \bar K^{*0}$ & 0.96 [0.76] & 0.96 [0.76] &
$\bar B^0\to K^0_1(1400) \bar K^{*0}$ & 0.67 [0.52] & 0.67 [0.52] \\
$\bar B^0\to \bar K^0_1(1270) K^{*0}$ & 0.0007 [-] & 0.31 [-] &
$\bar B^0\to \bar K^0_1(1400) K^{*0}$ & 1.1 [-] & 0.82 [-] \\
$B^-\to K^-_1(1270) K^{*0}$      & 0.0007 [-] & 0.33 [-] &
$B^-\to K^-_1(1400) K^{*0}$      & 1.2 [-] & 0.88 [-] \\
$B^-\to K^0_1(1270) K^{*-}$      & 1.0 [0.82] & 1.0 [0.82] &
$B^-\to K^0_1(1400) K^{*-}$      & 0.73 [0.56] & 0.73 [0.56] \\
\hline \hline
\end{tabular}
\end{center}
\end{table}

\begin{table}[ht]
{\small TABLE XIII. Branching ratios (in units of $10^{-6}$) of $B\to b_1(h_1) P$ and
$B\to b_1(h_1)V$ decays, calculated with decay constants in the limit $f_{b_1}=f_{h_1}=0$.}
\begin{center}
\begin{tabular}{l c}\hline \hline
Mode\, \, \, \,  & ${\cal B}$  \\
\hline
$\bar B^0\to b^-_1 \pi^+$  & 0.0 \\
$B^-\to b^-_1 \pi^0$       & 0.29 \\

$B^-\to b_1^- \eta$        & 0.061 \\
$B^-\to b_1^- \eta'$       & 0.58 \\

$\bar B^0\to b^-_1 \rho^+$ & 0.0 \\
$B^-\to b^-_1 \rho^0$      & 0.0005 \\

$\bar B^0\to b_1^0 \omega$ & 0.02 \\
$B^-\to b_1^- \omega$      & 0.004 \\

$\bar B^0\to h_1 \rho^0$   & 0.02 \\

$\bar B^0\to h_1 K^{*0}$   & 0.16 \\
$B^-\to h_1 K^{*-}$        & 0.34 \\

\hline \hline
\end{tabular}
\end{center}
\end{table}

\begin{table}[ht]
{\small TABLE XIV. Branching ratios (in units of $10^{-6}$) of $B\to AA$ decays,
where $A$ is a nonstrange axial-vector meson, using the ISGW2 form factors.}
\begin{center}
\begin{tabular}{l c}\hline \hline
Mode\, \, \, \,  & ${\cal B}$   \\
\hline
$\bar B^0\to a^-_1 a^+_1$ & 6.4  \\
$\bar B^0\to a^0_1 a^0_1$ & 0.1  \\
$B^-\to a^-_1 a^0_1$      & 3.6  \\
$\bar B^0\to a_1^0 f_1$   & 0.02 \\
$B^-\to a_1^- f_1     $   & 3.7  \\
\hline \hline
\end{tabular}
\end{center}
\end{table}

\begin{table}[ht]
{\small TABLE XV. Branching ratios (in units of $10^{-6}$) of $B\to K_1A$ decays,
where $A$ is a nonstrange axial-vector meson, using the ISGW2 form factors. The $K_1$
axial mesons are $K_1(1270)$ and $K_1(1400)$.}
\begin{center}
\begin{tabular}{l c c c c c }\hline \hline
Mode\, \, \, \, & ${\cal B}$ $(32^\circ)$ & ${\cal B}$ $(58^\circ)$
& Mode\, \, \, \,  & ${\cal B}$ $(32^\circ)$  & ${\cal B}$$(58^\circ)$\\
\hline
$\bar B^0\to K^-_1(1270) a^+_1$      & 0.79  & 0.79  &
$\bar B^0\to K^-_1(1400) a^+_1$      & 0.49  & 0.49   \\
$\bar B^0\to \bar K^0_1(1270) a^0_1$ & 0.002 & 0.03  &
$\bar B^0\to \bar K^0_1(1400) a^0_1$ & 0.08  & 0.06  \\
$B^-\to K^-_1(1270) a^0_1$           & 0.12  & 0.06  &
$B^-\to K^-_1(1400) a^0_1$           & 0.03  & 0.03  \\
$B^-\to \bar K^0_1(1270) a^-_1$      & 0.002 & 0.002 &
$B^-\to \bar K^0_1(1400) a^-_1$      & 0.001 & 0.001 \\

$\bar B^0\to \bar K_1^0(1270) f_1$  & 0.44 & 0.53 &
$\bar B^0\to \bar K_1^0(1400) f_1$  & 0.48 & 0.44 \\
$B^-\to K_1^-(1270) f_1$            & 0.15 & 0.27 &
$B^-\to K_1^-(1400) f_1$            & 0.34 & 0.29 \\
\hline \hline
\end{tabular}
\end{center}
\end{table}

\begin{table}[ht]
{\small TABLE XVI. Experimental bounds for branching ratios of some charmless
$B\to AP$, $AV$ and $AA$ decays reported in \cite{pdg2006}.}
\begin{center}
\begin{tabular}{l c}\hline \hline
Mode\, \, \, \,  & ${\cal B}^{exp}$   \\
\hline
$B^0\to a^{\mp}_1(1260) \pi^{\pm}$ & $<4.9\times10^{-4}$  \\
$B^0\to a^0_1(1260) \pi^0$ & $<1.1 \times10^{-3}$  \\
$B^0\to a^+_1(1260) \rho^-$      & $<3.4 \times10^{-3}$  \\
$B^0\to a_1^0(1260) \rho^0$   & $<2.4\times10^{-3}$ \\
$B^0\to K_1^+(1400) \pi^-$   & $<1.1\times10^{-3}$ \\
$B^0\to a_1^+(1260)K^-$   & $<2.3\times10^{-4}$ \\
$B^0\to K_1^0(1400) \rho^0$   & $<3.0\times10^{-3}$ \\
$B^0\to K_1^0(1400) \phi$   & $<5.0\times10^{-3}$ \\
$B^+\to a_1^+(1260) \pi^0$   & $<1.7\times10^{-3}$ \\
$B^+\to a_1^0(1260) \pi^+$   & $<9.0\times10^{-4}$ \\
$B^+\to a_1^+(1260) \rho^0$   & $<6.2\times10^{-4}$ \\
$B^+\to a_1^+(1260) a_1^0(1260)$   & $<1.3\%$ \\
$B^+\to K_1^0(1400) \pi^+$   & $<2.6\times10^{-3}$ \\
$B^+\to K_1^+(1400) \rho^0$   & $<7.8\times10^{-4}$ \\
$B^+\to K_1^+(1400) \phi$   & $<1.1\times10^{-3}$ \\
\hline \hline
\end{tabular}
\end{center}
\end{table}

\section{Conclusions}

In this work, we have presented a systematic study of exclusive charmless nonleptonic
two-body $B$ decays including axial-vector mesons in the final state. Branching ratios of
decays $B\to PA$, $B\to VA$ and $B\to AA$ (where $A$, $V$ and $P$ denote an axial-vector,
a vector and a pseudoscalar meson, respectively) have been calculated assuming the naive
factorization hypothesis and using the improved version of the nonrelativistic ISGW quark
model in order to obtain form factors required for $B\to A$ transitions. Form factors for
$B\to P$ and $B\to V$ transitions were obtained from the WSB model and LCSR approach. 
We have included contributions that arise from the effective $\Delta B=1$ weak Hamiltonian 
$H_{eff}$, i.e., we have considered $W$-external and $W$-internal emissions, which have 
contributions of $a_1$ and $a_2$ QCD parameters, respectively, and penguin contributions 
given by $a_{3,...,10}$ QCD parameters. The respective factorized amplitudes of these decays 
are explicitly showed in appendices and their penguin contributions have been classified. 
We also present a comparison between $B\to A$ and $B\to V$ transitions.

We have obtained branching ratios for 141 exclusive channels $B\to AP$, $AV$ and $AA$ where
the axial-vector meson can be a $^3P_1$ or a $^1P_1$ meson. We also studied the dependence of
the branching ratios for $B\to K_1P(V,A)$ decays ($K_1=K_1(1270),K_1(1400)$ are the physical
strange axial-vector mesons) with respect to the mixing angle between $K_A$ and $K_B$. The
best scenarios for determining this mixing angle are the decays
$\bar B^0\to \bar K_1^0(1270)\etap (K^0, K^{*0})$ and $B^-\to K_1^-(1270)\etap(K^0, K^{*0})$
because their branching ratios strongly depend on the mixing angle.

Our results show that some of these decays can be reach in experiment. In fact, decays 
$B^-\to a_1^0\pi^-$, $\bar B^0\to a_1^{\pm}\pi^{\mp}$, $B^-\to a_1^-\bar K^{0}$, 
$\bar B^0\to a_1^+K^-$, $\bar B^0\to f_1\bar K^0$, $B^-\to f_1K^-$, $B^-\to K_1^-(1400)\etap$, 
$B^-\to b_1^-\bar K^{0}$, and $\bar B^0\to b_1^+\pi^-(K^-)$ have branching ratios of the order 
of $10^{-5}$.

At experimental level there is not enough information. In Ref. \cite{pdg2006} there are only
bounds for branching ratios of some charmless $B \to AP$, $AV$, $AA$ decays (see Table XVI).
In general, our results are smaller that these bounds by two orders of magnitude. Our
predictions $Br(\bar B^0\to a_1^-\pi^+)=36.7\times 10^{-6}[23.5\times 10^{-6}]$ and
$Br(\bar B^0 \to a_1^+\pi^-)=74.3 \times 10^{-6}[-]$, i.e. the CP-averaged branching ratio 
$Br(\bar B^0\to a_1{^\mp} \pi^{\pm})=55.5\times 10^{-6}[48.9\times 10^{-6}]$
is consistent with the experimental average
value $Br(\bar B^0\to a_1^{\mp}\pi^{\pm})=(40.9\pm 7.6)\times 10^{-6}$ \cite{nardulli06}.
This average includes BaBar and Belle results \cite{exp}.

In general, we can explain the large branching ratios for $B\to K_1(1400)\etap$ as a 
combination of effects, the constructive interference of the terms $a_4$ and $a_6$ which 
are the bigger coefficients in the penguin sector of the effective Hamiltonian and the two
mixing $K_{1A}-K_{1B}$ and $\eta-\eta'$ involved in the decays.

Finally, we want to mention that our predictions are at the same order that ones obtained by
Laporta-Nardulli-Pham (see Tables V and VI in Ref. \cite{nardulli06}), although our values are
in general bigger, except in a few modes. On the other hand, predictions obtained in Ref.
\cite{chen05} (see Tables II and III) for $B\to K_1(1270)\phi$ and $B\to K_1(1400)\phi$ and
our predictions for these modes (see Table XI) do not agree, except for the case with
$N_c^{eff}=\infty$ and $\theta=58^\circ$ (with $\mu=2.5$ GeV or $\mu=4.4$ GeV). In this case
the respective branching ratios are ${\cal O}(10^{-7})$.

\begin{acknowledgments}
We thank C. Ram\'irez from Universidad Industrial de Santander, Colombia for useful 
conversatios and G. L\'opez Castro from CINVESTAV, M\'exico for reading the manuscript and 
his valuable suggestions. The authors acknowledge financial support from Conacyt (G. C.) and 
{\it Comit\'e Central de Investigaciones} of University of Tolima (J. H. M. and C. E. V.).
\end{acknowledgments}

\begin{appendix}

\section{Matrix elements for $B$ decays to an axial and a pseudoscalar meson}

\ba {\cal M}(\bar B^0\to a^-_1\pi^+) &=& 2 m_{a_1} f_{a_1}
F^{B\to \pi}_1(m^2_{a_1})
\Bigg\{V_{ub}V^*_{ud}\ a_1 - V_{tb}V^*_{td}(a_4+a_{10})\Bigg\}\ea

\ba {\cal M}(\bar B^0\to a^+_1\pi^-)  &=& 2i m_\pi f_\pi
V^{B\to a_1}_0(m^2_{\pi})\Bigg\{V_{ub}V^*_{ud}\ a_1\nn\\
&&-V_{tb}V^*_{td}\Bigg[a_4+a_{10}
+2(a_6+a_8)\frac{m^{2}_{\pi^-}}{(m_u+m_d)(m_b-m_u)}\Bigg] \Bigg\}\ea

\ba {\cal M}(\bar B^0\to a^0_1\pi^0) &=&2i m_\pi f_\pi
V^{B\to a_1}_0(m^2_\pi)\Bigg\{V_{ub}V^*_{ud}\ a_2\nn\\
&&-V_{tb}V^*_{td}\left[-a_4-(2a_6 - a_8
){m^2_\pi\over 2m_d(m_b-m_d)}-{1\over2}(3a_7-3a_9-a_{10})\right]\Bigg\}\nn\\
&&+2 m_{a_1} f_{a_1} F^{B\to \pi}_1 (m^2_{a_1})
\Bigg\{V_{ub}V^*_{ud}\ a_2-V_{ub}V^*_{ud}
\left[-a_4-{1\over 2}(3a_7-3a_9-a_{10})\right]\Bigg\} \ea

\ba {\cal M}(B^-\to a^-_1\pi^0) &=&  2i m_\pi f_\pi
V^{B\to a_1}_0(m^2_\pi)\Bigg\{V_{ub}V^*_{ud}a_2\nn\\
&& - V_{tb}V^*_{td}\left[-a_4-{1\over 2}(3a_7-3a_9-a_{10}) -(2
a_6-a_8){m^2_\pi \over 2m_d(m_b-m_d)}\right]\Bigg\}\nn\\
&& +2 m_{a_1} f_{a_1} F^{B\to \pi}_1 (m^2_{a_1})
\Bigg\{V_{ub}V^*_{ud}a_1- V_{tb}V^*_{td}(a_4+a_{10})\Bigg\}\ea

\ba {\cal M}(B^-\to a^0_1 \pi^-) &=& 2i m_\pi f_\pi V^{B\to a_1}_0(m^2_\pi)
\Bigg\{V_{ub}V^*_{ud}a_1-V_{tb}V^*_{td}\left[a_4+a_{10}+2(a_6+a_8){m^{2}_{\pi^-}
\over(m_d+m_u)(m_b-m_u)}\right]\Bigg\} \nn\\
&& +2 m_{a_1} f_{a_1} F^{B\to \pi}_1 (m_{a_1}^2)\Bigg\{ V_{ub}V^*_{ud}a_2
-V_{tb}V^*_{td}\left[-a_4-{1\over 2}(3a_7-3a_9-a_{10})\right]\Bigg\}\ea

\ba {\cal M}(\bar B^0\to a^0_1\etap) &=& 2i m_\etap f^u_\etap
V^{B\to a_1}_0(m^2_\etap)\Bigg\{V_{ub}V^*_{ud}\ a_2\nn\\ 
&&-V_{tb}V^*_{td}
\left[2a_3+a_4-2a_5+(2a_6-a_8){m^2_\etap \over 2m_s(m_b-m_d)}
\left({f^s_\etap \over f^u_\etap}-1\right)r_\etap -
{1\over 2}(a_7-a_9+a_{10})\right]\Bigg\}\nn\\
&&+2 m_{a_1} f_{a_1} F^{B\to \etap}_1(m^2_{a_1})\Bigg\{V_{ub}V^*_{ud}\ a_2
-V_{tb}V^*_{td}\left[-a_4+{1\over 2}a_{10}- \frac{3}{2}(a_7-a_9)\right]\Bigg\}\nn\\
&&-2i m_\etap f^s_\etap V^{B\to a_1}_0(m^2_\etap)\Bigg\{
V_{tb}V^*_{td}\left[a_3-a_5+\frac{1}{2}(a_7-a_9)\right]\Bigg\}\nn\\
&&+2i m_\etap f^c_\etap V^{B\to a_1}_0(m^2_\etap)\Bigg\{V_{cb}V^*_{cd}\ a_2
-V_{tb}V^*_{td}\left(a_3-a_5-a_7+a_9\right)\Bigg\}\ea

\ba {\cal M}(B^-\to a^-_1\etap) &=& 2i m_\etap f^u_\etap 
V^{B\to a_1}_0(m^2_\etap)\Bigg\{V_{ub}V^*_{ud} a_2\nn\\
&&-V_{tb}V^*_{td}\left[2a_3+a_4-2a_5-\frac{1}{2}(a_7-a_9+a_{10})+(2a_6-a_8)
{m^2_\etap \over 2m_s(m_b-m_d)}\left({f^s_\etap \over f^u_\etap}-1\right)
r_\etap\right]\Bigg\}\nn\\
&&+2 m_{a_1} f_{a_1} F^{B\to \etap}_1(m^2_{a_1})\Bigg\{V_{ub}V^*_{ud} a_1
-V_{tb}V^*_{td}\left(a_4+a_{10}\right)\Bigg\}\nn\\
&&-2i m_\etap f^s_\etap V^{B\to a_1}_0(m^2_\etap)\Bigg\{
V_{tb}V^*_{td}\left[2a_3+a_4-2a_5-\frac{1}{2}(a_7-a_9)\right]\Bigg\}\nn\\
&& +2i m_\etap f^c_\etap V^{B\to a_1}_0(m^2_\etap)\Bigg\{V_{cb}V^*_{cd}\ a_2 -
V_{tb}V^*_{td}\left(a_3-a_5-a_7+a_9 \right)\Bigg\}\ea

\ba {\cal M}(\bar B^0 \to a^0_1 \bar K^0) &=& 2 m_{a_1} f_{a_1}
F^{B\to K}_1 (m_{a_1}^2)\Bigg\{V_{ub}V^*_{us}\ a_2- V_{tb}V^*_{ts}
{3\over 2}(-a_7+a_9)\Bigg\}\nn\\
&&-2i m_K f_K V^{B\to a_1}_0(m^2_K)V_{tb}V^*_{ts}
\left[a_4-{1\over 2}a_{10}+(2a_6-a_8){m^2_K \over(m_s+m_d)(m_b-m_d)}\right]\ea

\ba {\cal M}(\bar B^0\to a^+_1 K^-) &=& 2i m_K f_K V^{B\to a_1}_0(m^2_K)
\Bigg\{ V_{ub}V^*_{us}\ a_1\nn\\
&& -V_{tb}V^*_{ts}\left[a_4+a_{10}+
2(a_6+a_8){m^2_K\over(m_s+m_u)(m_b-m_u)}\right]\Bigg\}\ea

\ba {\cal M}(B^-\to a^0_1 K^-) &=& 2i m_K f_K V^{B\to a_1}_0(m^2_K)
\Bigg\{V_{ub}V^*_{us} a_1 - V_{tb}V^*_{ts}
\left[a_4+a_{10}+2(a_6+a_8){m^{2}_{K^-}\over(m_s+m_u)(m_b-m_u)}\right]\Bigg\}\nn\\
&& +2 m_{a_1} f_{a_1} F^{B\to K}_1(m^2_{a_1})(\epsilon^* \cdot p_B)
\Bigg\{V_{ub}V^*_{us}a_2-V_{tb}V^*_{ts}{3\over2}(-a_7+a_9)\Bigg\}\ea

\ba {\cal M}(B^- \to a^-_1 \bar K^0) &=& -2i m_K f_K 
V^{B\to a_1}_0(m^2_K)V_{tb}V^*_{ts} 
\left[a_4-{1\over 2}a_{10}+(2a_6-a_8){m^2_K\over(m_s+m_d)(m_b-m_d)}\right]\ea

\ba {\cal M}(\bar B^0 \to f_1\pi^0) &=& 2i m_\pi f_\pi
V^{B\to f_1}_0(m^2_\pi)\Bigg\{V_{ub}V^*_{ud}\ 
a_2 -V_{tb}V^*_{td}\left[a_4-{1\over 2}(3a_7-3a_9+a_{10})
+(2a_6-a_8){m^{2}_{\pi^0}\over 2m_d(m_b-m_d)}\right]\Bigg\}\nn\\
&& +2 m_{f_1} f_{f_1} F^{B\to \pi}_1(m_{f_1}^2)
\Bigg\{V_{ub}V^*_{ud}\ a_2-V_{tb}V^*_{td}\left[2a_3+a_4-2a_5-{
1\over 2}(a_7-a_9+a_{10})\right]\Bigg\}\ea

\ba {\cal M}(B^- \to f_1\pi^-) &=& 2i m_{\pi} f_{\pi} V^{B\to f_1}_0(m^2_\pi)
\Bigg\{V_{ub}V^*_{ud} a_1
-V_{tb}V^*_{td}\left[a_4+a_{10}+2(a_6+a_8){m^{2}_{\pi^-}
\over(m_d+m_u)(m_b-m_u)}\right]\Bigg\}\nn\\
&& +2 m_{f_1} f_{f_1} F^{B\to \pi}_1(m^2_{f_1})
\Bigg\{V_{ub}V^*_{ud} a_2 -V_{tb}V^*_{td}
\left[2a_3+a_4-2a_5-{1\over2} (a_7-a_9+a_{10})\right]\Bigg\}\ea

\ba {\cal M}(\bar B^0 \to f_1\etap) &=& 2i m_\etap f^u_\etap
V^{B\to f_1}_0(m^2_\etap)\Bigg\{V_{ub}V^*_{ud}\ a_2\nn\\
&&-V_{tb}V^*_{td}\left[2a_3+a_4-2a_5-{1\over2}(a_7-a_9+a_{10})+(2a_6-a_8)
{m^2_\etap\over 2m_d(m_b-m_d)}\left({f^s_\etap \over f^u_\etap}-1\right)
r_\etap\right]\Bigg\}\nn\\
&& +2 m_{f_1} f_{f_1} F^{B\to \etap}_1(m^2_{f_1})\Bigg\{ V_{ub}V^*_{ud}\ a_2
-V_{tb}V^*_{td}\left[2a_3+a_4-2a_5-{1\over2}(a_7-a_9+a_{10})\right]\Bigg\}\nn\\
&&-2i m_\etap f^s_\etap V^{B\to f_1}_0(m^2_\etap)\Bigg\{V_{tb}V^*_{td}
\left[a_3-a_5+{1\over2}(a_7-a_9)\right]\Bigg\}\nn\\
&&+2i m_\etap f^c_\etap V^{B\to f_1}_0(m^2_\etap)\Bigg\{V_{cb}V^*_{cd}\ a_2
-V_{tb}V^*_{td}\left[a_3-a_5-a_7+a_9\right]\Bigg\}\ea

\ba {\cal M}(\bar B^0 \to f_1 \bar K^0) &=&  2 m_{f_1} f_{f_1}
F^{B\to K}_1(m^2_{f_1})\Bigg\{V_{ub}V^*_{us}\
a_2 - V_{tb}V^*_{ts} \left[2a_3-2a_5-{1\over2}(a_7-a_9)\right]\Bigg\}\nn\\
&& -2i m_K f_K V^{B\to f_1}_0(m^2_K)V_{tb}V^*_{ts}
\left[a_4-{1\over 2}a_{10}+(2a_6-a_8){m^2_K \over(m_s+m_d)(m_b-m_d)}\right]\ea

\ba {\cal M}(B^-\to f_1 K^-) &=& 2i m_K f_K V^{B\to f_1}_0(m^2_K)
\Bigg\{V_{ub}V^*_{us} a_1-V_{tb}V^*_{ts}
\left[a_4+a_{10}+2(a_6+a_8){m^{2}_{K^-}\over(m_s+m_u)(m_b-m_u)}\right]\Bigg\}\nn\\
&& +2 m_{f_1} f_{f_1} F^{B\to K}_1(m^2_{f_1})\Bigg\{V_{ub}V^*_{us} a_2
-V_{tb}V^*_{ts}\left[2a_3-2a_5-{1\over 2}(a_7-a_9)\right] \Bigg\}\ea

\ba {\cal M}(\bar B^0 \to K^-_1\pi^+) &=& 2i m_\pi f_\pi 
V^{B\to K_1}_0(m^2_\pi)
\Bigg\{V_{ub}V^*_{us}\ a_1 - V_{tb}V^*_{ts}\left(a_4+a_{10}\right)\Bigg\}\ea

\ba {\cal M}(\bar B^0\to \bar K^0_1\pi^0) &=& 2i m_\pi f_\pi 
V^{B\to K^0_1}_0(m^2_\pi)
\Bigg\{V_{ub}V^*_{us}\ a_2 - V_{tb}V^*_{ts}{3\over 2}(-a_7+a_9)\Bigg\}\nn\\
&& -2 m_{K_1} f_{K_1} F^{B\to \pi}_1(m^2_{K_1})
V_{tb}V^*_{ts}\left[a_4-{1\over 2}a_{10}\right]\ea

\ba {\cal M}(B^-\to K^-_1\pi^0) &=& 2i m_\pi f_\pi
V^{B\to K_1}_0(m^2_\pi)\Bigg\{V_{ub}V^*_{us}a_2
-V_{tb}V^*_{ts}{3\over 2}(-a_7+a_9)\Bigg\}\nn\\
&&+2 m_{K_1} f_{K_1} F^{B\to \pi}_1(m^2_{K_1(1270)})
\Bigg\{V_{ub}V^*_{us}a_1-V_{tb}V^*_{ts}(a_4+a_{10})\Bigg\}\ea

\ba {\cal M}(B^-\to \bar K^0_1\pi^-) &=& -2 m_{K_1} f_{K_1} 
F^{B\to \pi}_1(m^2_{K_1})V_{tb}V^*_{ts}
\Bigg\{\left(a_4-{1\over 2}a_{10}\right) \Bigg\}\ea

\ba {\cal M}(\bar B^0 \to \bar K^0_1\etap) &=& 2i m_\etap f^u_\etap 
V^{B\to K_1}_0(m^2_\etap)\Bigg\{V_{ub}V^*_{us}\ a_2 -V_{tb}V^*_{ts}
\left[2(a_3-a_5)-{1\over 2}(a_7-a_9)\right]\Bigg\}\nn\\
&&-2 m_{K_1} f_{K^0_1} F^{B\to \etap}_1(m^2_{K^0_1})
\Bigg\{V_{tb}V^*_{ts}\left(a_4-{1\over 2}a_{10}\right)\Bigg\}\nn\\
&&-2i m_\etap f^s_\etap V^{B\to K_1}_0(m^2_\etap)\Bigg\{
V_{tb}V^*_{ts}\nn\\
&&\left[a_3+a_4-a_5+{1\over 2}(a_7-a_9-a_{10})
+(2a_6-a_8){m^2_\etap\over 2m_s(m_b-m_s)}
\left(1-{f^u_\etap \over f^s_\etap}\right)\right]\Bigg\}\nn\\
&&+2i m_\etap f^c_\etap V^{B\to K_1}_0(m^2_\etap)
\Bigg\{V_{ub}V^*_{us}\ a_2-V_{tb}V^*_{ts}(a_3-a_5-a_7+a_9)\Bigg\}\ea

\ba{\cal M}(B^- \to K^-_1\etap) &=& 2i m_\etap f^u_\etap
V^{B\to K_1}_0(m^2_\etap)\Bigg\{V_{ub}V^*_{us}\ a_2 - 
V_{tb}V^*_{ts}\left[2a_3-2a_5-{1\over 2}(a_7-a_9)\right]\Bigg\}\nn\\
&&+2 m_{K_1} f_{K_1} F^{B\to \etap}_1(m^2_{K_1})
\Bigg\{V_{ub}V^*_{us}\ a_1 -V_{tb}V^*_{ts}(a_4+a_{10})\Bigg\}\nn\\
&&-2i m_\etap f^s_\etap V^{B\to K_1}_0(m^2_\etap)
\Bigg\{V_{tb}V^*_{ts}\nn\\
&&\left[a_3+a_4-a_5+{1\over 2}(a_7-a_9-a_{10})
+(2a_6-a_8){m^2_\etap\over 2m_s(m_b-m_s)}
\left(1-{f^u_\etap \over f^s_\etap}\right)\right]\Bigg\}\nn\\
&&+2i m_\etap f^c_\etap V^{B\to K_1}_0(m^2_\etap)
\Bigg\{V_{cb}V^*_{cs}\ a_2-V_{tb}V^*_{ts}(a_3-a_5-a_7+a_9)\Bigg\} \ea

\ba {\cal M}(\bar B^0 \to K^0_1\bar K^0) &=& -2 m_{K_1} f_{K_1} 
F^{B\to K}_1(m^2_{K_1})V_{tb}V^*_{td}
\Bigg\{\left(a_4-{1\over 2}a_{10}\right)\Bigg\}\ea

\ba{\cal M}(\bar B^0 \to\bar K^0_1 K^0) &=& -2i m_K f_K 
V^{B\to K_1}_0(m^2_K)V_{tb}V^*_{td}\Bigg\{\left[a_4-{1\over 2}
a_{10}+(2a_6-a_8){m^2_K \over(m_s+m_d)(m_b-m_s)}\right]\Bigg\}\ea

\ba {\cal M}(B^-\to K^-_1K^0) &=& -2i m_K f_K
V^{B\to K_1}_0(m^2_K)V_{tb}V^*_{td} \Bigg\{\left[a_4-{1\over 2}a_{10}
+(2a_6-a_8){m^{2}_{K^0}\over(m_s+m_d)(m_b-m_s)}\right] \Bigg\}\ea

\ba {\cal M}(B^-\to K^0_1  K^-) &=& -2 m_{K_1} f_{K_1} 
F^{B\to K}_1 (m^2_{K_1})V_{tb}V^*_{td} 
\Bigg\{\left(a_4-{1\over 2}a_{10}\right)\Bigg\}\ea

\section{Matrix elements for $B$ decays to an axial and a vector meson}

\ba {\cal M}(\bar B^0 \to a^-_1 \rho^+) &=& m_{a_1} f_{a_1}
\Bigg\{V_{ub}V^*_{ud}\ a_1 - V_{tb}V^*_{td}(a_4+a_{10})\Bigg\}
\Bigg(\frac{2V^{B\to \rho}(m_{a_1} ^2)}{(m_B+m_\rho)}
\epsilon_{\mu\nu\alpha\beta}\epsilon^\mu_{a_1}\epsilon^\nu_\rho p^\alpha_B p^\beta_\rho\nn\\
&&-i(m_B+m_\rho)A^{B\to \rho}_1(m^2_{a_1})(\epsilon_\rho
\cdot \epsilon_{a_1})+\frac{iA^{B\to \rho}_2(m^2_{a_1})}{(m_B+m_\rho)}
(\epsilon_\rho \cdot p_B)(\epsilon_{a_1} \cdot p_B)\Bigg)\ea

\ba {\cal M}(\bar B^0 \to a^+_1 \rho^-) &=& -m_\rho f_\rho
\Bigg\{V_{ub}V^*_{ud}\ a_1-V_{tb}V^*_{td}(a_4+a_{10})\Bigg\}
\Bigg(\frac{2A^{B\to a_1}(m^2_\rho)}{(m_B+m_{a_1})}
\epsilon_{\mu\nu\alpha\beta}\epsilon^\mu_\rho\epsilon^\nu_{a_1} p^\alpha_B p^\beta_{a_1}\nn\\ 
&&-i(m_B+m_{a_1})V^{B\to a_1}_1(m^2_\rho)(\epsilon_{a_1} \cdot \epsilon_\rho) 
+ \frac{iV^{B\to a_1}_2(m^2_\rho)}{(m_B+m_{a_1})}(\epsilon_{a_1}\cdot p_B)
(\epsilon_\rho \cdot p_B)\Bigg)\ea

\ba {\cal M}(\bar B^0 \to a^0_1\rho^0) &=& -m_\rho f_\rho
\Bigg\{V_{ub}V^*_{ud}\ a_2- V_{tb}V^*_{td}
\left[-a_4+\frac{1}{2}(3a_7+3a_9+a_{10})\right]\Bigg\}
\Bigg(\frac{2A^{B\to a_1}(m^2_\rho)}{(m_B+m_{a_1})}
\epsilon_{\mu\nu\alpha\beta}\epsilon^\mu_\rho\epsilon^\nu_{a_1}
p^\alpha_B p^\beta_{a_1}\nn\\
&&-i(m_B+m_{a_1})V^{B\to a_1}_1(m^2_\rho)(\epsilon_{a_1} \cdot \epsilon_\rho) 
+ \frac{iV^{B\to a_1}_2(m^2_\rho)}{(m_B+m_{a_1})}(\epsilon_{a_1} \cdot p_B)
(\epsilon_\rho \cdot p_B)\Bigg)\nn\\
&&+m_{a_1} f_{a_1} \Bigg\{V_{ub}V^*_{ud}\ a_2- V_{tb}V^*_{td}
\left[-a_4-\frac{1}{2}(3a_7-3a_9-a_{10})\right]\Bigg\}
\Bigg(\frac{2V^{B\to \rho}(m^2_{a_1})}{(m_B+m_\rho)} 
\epsilon_{\mu\nu\alpha\beta}\epsilon^\mu_{a_1}\epsilon^\nu_\rho
p^\alpha_B p^\beta_\rho\nn\\ 
&&- i(m_B+m_\rho)A^{B\to \rho}_1(m^2_{a_1})(\epsilon_\rho \cdot \epsilon_{a_1}) + 
\frac{iA^{B\to \rho}_2(m^2_{a_1})}{(m_B+m_{\rho})}(\epsilon_{\rho}\cdot p_B)
(\epsilon_{a_1} \cdot p_B)\Bigg)\ea

\ba {\cal M}(B^- \to a^-_1 \rho^0) &=&-m_\rho f_\rho
\Big\{V_{ub}V^*_{ud}a_2\Big\}
\Bigg(\frac{2A^{B\to a_1}(m^2_\rho)}{(m_B+m_{a_1})}
\epsilon_{\mu\nu\alpha\beta}\epsilon^\mu_\rho\epsilon^\nu_{a_1}
p^\alpha_B p^\beta_{a_1}\nn\\
&&-i(m_B+m_{a_1})V^{B\to a_1}_1(m^2_\rho)(\epsilon_{a_1} \cdot \epsilon_\rho) 
+\frac{iV^{B\to a_1}_2(m^2_\rho)}{(m_B+m_{a_1})} (\epsilon_{a_1} \cdot p_B)
(\epsilon_\rho \cdot p_B)\Bigg)\nn\\
&&+m_{a_1} f_{a_1} 
\Bigg\{V_{ub}V^*_{ud} a_1 -V_{tb}V^*_{td}\frac{3}{2}(-a_7+a_9+a_{10})\Bigg\}
\Bigg(\frac{2V^{B\to \rho}(m^2_{a_1})}{(m_B+m_\rho)} 
\epsilon_{\mu\nu\alpha\beta}\epsilon^\mu_{a_1}\epsilon^\nu_\rho
p^\alpha_B p^\beta_\rho\nn\\
&&-i(m_B+m_\rho)A^{B\to \rho}_1(m^2_{a_1})(\epsilon_\rho \cdot \epsilon_{a_1}) + 
\frac{iA^{B\to \rho}_2(m^2_{a_1})}{(m_B+m_\rho)}(\epsilon_\rho \cdot p_B)
(\epsilon_{a_1} \cdot p_B)\Bigg)\ea

\ba {\cal M}(B^- \to a^0_1 \rho^-) &=& -m_\rho f_\rho
\Bigg\{V_{ub}V^*_{ud} a_1-V_{tb}V^*_{td}\frac{3}{2}(a_7+a_9+a_{10})\Bigg\}
\Bigg(\frac{2A^{B\to a_1}(m^2_\rho)}{(m_B+m_{a_1})}
\epsilon_{\mu\nu\alpha\beta}\epsilon^\mu_\rho\epsilon^\nu_{a_1}
p^\alpha_B p^\beta_{a_1}\nn\\
&&-i(m_B+m_{a_1})V^{B\to a_1}_1(m^2_\rho)(\epsilon_{a_1} \cdot \epsilon_\rho) + 
\frac{iV^{B\to a_1}_2(m^2_\rho)}{(m_B+m_{a_1})}(\epsilon_{a_1} \cdot p_B)
(\epsilon_\rho \cdot p_B)\Bigg)\nn\\
&&+m_{a_1} f_{a_1} V_{ub}V^*_{ud}a_2
\Bigg(\frac{2V^{B\to \rho}(m^2_{a_1})}{(m_B+m_{\rho})} 
\epsilon_{\mu\nu\alpha\beta}\epsilon^\mu_{a_1}\epsilon^\nu_\rho
p^\alpha_B p^\beta_\rho\nn\\
&&-i(m_B+m_\rho)A^{B\to \rho}_1(m^2_{a_1})(\epsilon_\rho \cdot \epsilon_{a_1}) + 
\frac{iA^{B\to \rho}_2(m^2_{a_1})}{(m_B+m_\rho)}(\epsilon_\rho \cdot p_B)
(\epsilon_{a_1} \cdot p_B)\Bigg)\ea

\ba {\cal M}(\bar B^0 \to a^+_1 K^{*-}) &=& -m_{K^*} f_{K^*}
\Bigg\{V_{ub}V^*_{us}\ a_1 - V_{tb}V^*_{ts}(a_4+a_{10})\Bigg\}
\Bigg(\frac{2A^{B\to a_1}(m^2_{K^*})}{(m_B+m_{a_1})}
\epsilon_{\mu\nu\alpha\beta}\epsilon^\mu_{K^*}\epsilon^\nu_{a_1}
p^\alpha_B p^\beta_{a_1}\nn\\ 
&&-i(m_B+m_{a_1})V^{B\to a_1}_1(m^2_{K^*})(\epsilon_{a_1} \cdot \epsilon_{K^*}) 
+ \frac{iV^{B\to a_1}_2(m^2_{K^*})}{(m_B+m_{a_1})}(\epsilon_{a_1} \cdot p_B)
(\epsilon_{K^*}\cdot p_B)\Bigg)\ea

\ba {\cal M}(\bar B^0 \to a^0_1 \bar K^{*0}) &=&m_{K^*}f_{K^*} 
V_{tb}V^*_{ts}\Bigg\{\left(a_4-\frac{1}{2}a_{10}\right)\Bigg\}
\Bigg(\frac{2A^{B\to a_1}(m^2_{K^*})}{(m_B+m_{a_1})} 
\epsilon_{\mu\nu\alpha\beta}\epsilon^\mu_{K^*}\epsilon^\nu_{a_1}
p^\alpha_B p^\beta_{a_1}\nn\\ 
&&-i(m_B+m_{a_1})V^{B\to a_1}_1(m^2_{K^*})(\epsilon_{a_1} \cdot \epsilon_{K^*}) + 
\frac{iV^{B\to a_1}_2(m^2_{K^*})}{(m_B+m_{a_1})}(\epsilon_{a_1}\cdot p_B)
(\epsilon_{K^*} \cdot p_B)\Bigg)\nn\\
&&+ m_{a_1} f_{a_1} 
\Bigg\{V_{ub}V^*_{us}\ a_2 +V_{tb}V^*_{ts}\frac{3}{2}(a_7-a_9)\Bigg\}
\Bigg(\frac{2V^{B\to K^*}(m^2_{a_1})}{(m_B+m_{K^*})} 
\epsilon_{\mu\nu\alpha\beta}\epsilon^\mu_{a_1}\epsilon^\nu_{K^*}
p^\alpha_B p^\beta_{K^*}\nn\\
&&-i(m_B+m_{K^*}A^{B\to K^*}_1(m^2_{a_1})(\epsilon_{K^*} \cdot \epsilon_{a_1}) + 
\frac{iA^{B\to K^*}_2(m^2_{a_1})}{(m_B+m_{K^*})}(\epsilon_{K^*} \cdot p_B)
(\epsilon_{a_1} \cdot p_B)\Bigg)\ea

\ba {\cal M}(B^- \to a^0_1 K^{*-}) &=& -m_{K^*} f_{K^*}
\Bigg\{V_{ub}V^*_{us} a_1 - V_{tb}V^*_{ts}(a_4+a_{10})\Bigg\}
\Bigg(\frac{2A^{B\to a_1}(m^2_{K^*})}{(m_B+m_{a_1})}
\epsilon_{\mu\nu\alpha\beta}\epsilon^\mu_{K^*}\epsilon^\nu_{a_1}
p^\alpha_B p^\beta_{a_1}\nn\\
&&-i(m_B+m_{a_1})V^{B\to a_1}_1(m^2_{K^*})(\epsilon_{a_1} \cdot \epsilon_{K^*}) 
+\frac{iV^{B\to a_1}_2(m^2_{K^*})}{(m_B+m_{a_1})}(\epsilon_{a_1} \cdot p_B)
(\epsilon_{K^*} \cdot p_B)\Bigg)\nn\\
&&+m_{a_1} f_{a_1}
\Bigg\{V_{ub}V^*_{us}a_2 +V_{tb}V^*_{ts}\frac{3}{2}(a_7-a_9)\Bigg\}
\Bigg(\frac{2V^{B\to {K^*}}(m^2_{a_1})}{(m_B+m_{K^*})} 
\epsilon_{\mu\nu\alpha\beta}\epsilon^\mu_{a_1}\epsilon^\nu_{K^*}
p^\alpha_B p^\beta_{K^*}\nn\\
&&-i(m_B+m_{K^*})A^{B\to K^*}_1(m^2_{a_1})(\epsilon_{K^*} \cdot \epsilon_{a_1})+
\frac{iA^{B\to K^*}_2(m^2_{a_1})}{(m_B+m_{K^*})}(\epsilon_{K^*} \cdot p_B)
(\epsilon_{a_1} \cdot p_B)\Bigg)\ea

\ba {\cal M}(B^- \to a^-_1\bar K^{*0}) &=& m_{K^*} f_{K^*} 
\Bigg\{V_{tb}V^*_{ts}\left(a_4-\frac{1}{2}a_{10}\right)\Bigg\}
\Bigg(\frac{2A^{B\to a_1}(m^2_{K^*})}{(m_B+m_{a_1})}
\epsilon_{\mu\nu\alpha\beta}\epsilon^\mu_{K^*}\epsilon^\nu_{a_1}
p^\alpha_B p^\beta_{a_1}\nn\\
&&-i(m_B+m_{a_1})V^{B\to a_1}_1(m^2_{K^*})
(\epsilon_{a_1}\cdot \epsilon_{K^*})+\frac{iV^{B\to a_1}_2(m^2_{K^*})}
{(m_B+m_{a_1})}(\epsilon_{a_1}\cdot p_B)(\epsilon_{K^*}\cdot p_B)\Bigg)\ea

\ba {\cal M}(\bar B^0 \to a^0_1\omega) &=& -m_\omega f_\omega
\Bigg\{V_{ub}V^*_{ud}\ a_2 -V_{tb}V^*_{td}\left[2a_3+a_4+2a_5+\frac{1}{2}
(a_7+a_9-a_{10})\right]\Bigg\}
\Bigg(\frac{2A^{B\to a_1}(m^2_\omega)}{(m_B+m_{a_1})}
\epsilon_{\mu\nu\alpha\beta}\epsilon^\mu_\omega\epsilon^\nu_{a_1}
p^\alpha_B p^\beta_{a_1}\nn\\
&&-i(m_B+m_{a_1})V^{B\to a_1}_1(m^2_{\omega})(\epsilon_{a_1}\cdot \epsilon_\omega) 
+ \frac{iV^{B\to a_1}_2(m^2_\omega)}{(m_B+m_{a_1})}(\epsilon_{a_1} \cdot p_B)
(\epsilon_\omega \cdot p_B)\Bigg)\nn\\
&& + m_{a_1} f_{a_1} 
\Bigg\{V_{ub}V^*_{ud}\ a_2 - V_{tb}V^*_{td}
\left[-a_4-\frac{1}{2}(3a_7-3a_9-a_{10})\right]\Bigg\}
\Bigg(\frac{2V^{B\to \omega}(m^2_{a_1})}{(m_B+m_\omega)} 
\epsilon_{\mu\nu\alpha\beta}\epsilon^\mu_{a_1}\epsilon^\nu_\omega
p^\alpha_B p^\beta_\omega\nn\\
&&- i(m_B+m_\omega)A^{B\to \omega}_1(m^2_{a_1})(\epsilon_\omega \cdot \epsilon_{a_1}) + 
\frac{iA^{B\to \omega}_2(m^2_{a_1})}{(m_B+m_\omega)}(\epsilon_\omega \cdot p_B)
(\epsilon_{a_1}\cdot p_B)\Bigg)\ea

\ba {\cal M}(B^- \to a^-_1 \omega) &=& -m_\omega f_\omega
\Bigg\{V_{ub}V^*_{ud} a_2-V_{tb}V^*_{td}\left[2a_3+a_4+2a_5+
\frac{1}{2}(a_7+a_9-a_{10})\right]\Bigg\}
\Bigg(\frac{2A^{B\to a_1}(m^2_\omega)}{(m_B+m_{a_1})}
\epsilon_{\mu\nu\alpha\beta}\epsilon^\mu_\omega\epsilon^\nu_{a_1}
p^\alpha_B p^\beta_{a_1}\nn\\
&&-i(m_B+m_{a_1})V^{B\to a_1}_1(m^2_\omega)(\epsilon_{a_1} \cdot \epsilon_\omega) 
+ \frac{iV^{B\to a_1}_2(m^2_\omega)}{(m_B+m_{a_1})}(\epsilon_{a_1}\cdot p_B)
(\epsilon_\omega \cdot p_B)\Bigg)\nn\\
&& +m_{a_1} f_{a_1}
\Bigg\{V_{ub}V^*_{ud}a_1 - V_{tb}V^*_{td}(a_4+a_{10})\Bigg\}
\Bigg(\frac{2V^{B\to \omega}(m^2_{a_1})}{(m_B+m_\omega)} 
\epsilon_{\mu\nu\alpha\beta}\epsilon^\mu_{a_1}\epsilon^\nu_\omega
p^\alpha_B p^\beta_\omega\nn\\
&&-i(m_B+m_\omega)A^{B\to \omega}_1(m^2_{a_1})(\epsilon_\omega \cdot \epsilon_{a_1}) + 
\frac{iA^{B\to \omega}_2(m^2_{a_1})}{(m_B+m_\omega)}(\epsilon_\omega \cdot p_B)
(\epsilon_{a_1}\cdot p_B)\Bigg)\ea

\ba {\cal M}(\bar B^0 \to a^0_1\phi) &=& m_\phi f_\phi
V_{tb}V^*_{td}\Bigg\{\left[a_3+a_5-\frac{1}{2}(a_7+a_9)\right]\Bigg\}
\Bigg(\frac{2A^{B\to a_1}(m^2_\phi)}{(m_B+m_{a_1})}
\epsilon_{\mu\nu\alpha\beta}\epsilon^\mu_\phi\epsilon^\nu_{a_1}
p^\alpha_B p^\beta_{a_1}\nn\\
&&-i(m_B+m_{a_1})V^{B\to a_1}_1(m^2_\phi)(\epsilon_{a_1}\cdot \epsilon_\phi) 
+ \frac{iV^{B\to a_1}_2(m^2_\phi)}{(m_B+m_{a_1})}(\epsilon_{a_1} \cdot p_B)
(\epsilon_\phi \cdot p_B)\Bigg)\ea

\ba {\cal M}(B^-\to a^-_1\phi) &=& -\sqrt{2}{\cal M}(\bar B^0\to a^0_1\phi)\ea

\ba {\cal M}(\bar B^0 \to f_1 \rho^0) &=& -m_\rho f_\rho
\Bigg\{V_{ub} V^*_{ud}\ a_2
-V_{tb}V^*_{td}\left[-a_4+\frac{1}{2}(3a_7+3a_9+a_{10})\right]\Bigg\}
\Bigg(\frac{2A^{B\to f_1}(m^2_\rho)}{(m_B+m_{f_1})}
\epsilon_{\mu\nu\alpha\beta}\epsilon^\mu_\rho\epsilon^\nu_{f_1}
p^\alpha_B p^\beta_{f_1}\nn\\
&&-i(m_B+m_{f_1})V^{B\to f_1}_1(m^2_\rho)(\epsilon_{f_1} \cdot \epsilon_\rho) 
+ \frac{iV^{B\to f_1}_2(m^2_\rho)}{(m_B+m_{f_1})}(\epsilon_{f_1} \cdot p_B)
(\epsilon_\rho \cdot p_B)\Bigg)\nn\\
&& +m_{f_1} f_{f_1} 
\Bigg\{V_{ub} V^*_{ud}\ a_2 -V_{tb}V^*_{td}
\left[2a_3+a_4-2a_5-\frac{1}{2}(a_7-a_9+a_{10})\right]\Bigg\}
\Bigg(\frac{2V^{B\to \rho}(m^2_{f_1})}{(m_B+m_\rho)}
\epsilon_{\mu\nu\alpha\beta}\epsilon^\mu_{f_1}\epsilon^\nu_\rho
p^\alpha_B p^\beta_\rho\nn\\
&&-i(m_B+m_\rho)A^{B\to \rho}_1(m^2_{f_1})(\epsilon_\rho \cdot \epsilon_{f_1}) + 
\frac{iA^{B\to \rho}_2(m^2_{f_1})}{(m_B+m_\rho)}(\epsilon_\rho \cdot p_B)
(\epsilon_{f_1} \cdot p_B)\Bigg)\ea

\ba {\cal M}(B^- \to f_1 \rho^-) &=& -m_\rho^- f_\rho
\Bigg\{V_{ub}V^*_{ud} a_1-V_{tb}V^*_{td} \left(a_4+a_{10}\right)\Bigg\}
\Bigg(\frac{2A^{B\to f_1}(m^2_\rho)}{(m_B+m_{f_1})}
\epsilon_{\mu\nu\alpha\beta}\epsilon^\mu_\rho\epsilon^\nu_{f_1}
p^\alpha_B p^\beta_{f_1}\nn\\
&&-i(m_B+m_{f_1})V^{B\to f_1}_1(m^2_{\rho^-})(\epsilon_{f_1} \cdot \epsilon_\rho) 
+ \frac{iV^{B\to f_1}_2(m^2_\rho)}{(m_B+m_{f_1})} (\epsilon_{f_1} \cdot p_B)
(\epsilon_\rho \cdot p_B)\Bigg)\nn\\
&& +m_{f_1} f_{f_1} 
\Bigg\{V_{ub}V^*_{ud} a_2- V_{tb}V^*_{td}
\left[2a_3+a_4-2a_5-\frac{1}{2}(a_7-a_9+a_{10})\right]\Bigg\}
\Bigg(\frac{2V^{B\to \rho}(m^2_{f_1})}{(m_B+m_\rho)} 
\epsilon_{\mu\nu\alpha\beta}\epsilon^\mu_{f_1}\epsilon^\nu_\rho
p^\alpha_B p^\beta_\rho\nn\\
&&-i(m_B+m_\rho)A^{B\to \rho}_1(m^2_{f_1})(\epsilon_\rho \cdot \epsilon_{f_1}) + 
\frac{iA^{B\to \rho}_2(m^2_{f_1})}{(m_B+m_\rho)}(\epsilon_\rho \cdot p_B)
(\epsilon_{f_1} \cdot p_B)\Bigg)\ea

\ba {\cal M}(\bar B^0 \to f_1 \bar K^{*0}) &=& m_{K^*} f_{K^*} 
V_{tb}V^*_{ts}\Bigg\{\left(a_4-\frac{1}{2}a_{10}\right)\Bigg\}
\Bigg(\frac{2A^{B\to f_1}(m^2_{K^*})}{(m_B+m_{f_1})}
\epsilon_{\mu\nu\alpha\beta}\epsilon^\mu_{K^*}\epsilon^\nu_{f_1}
p^\alpha_B p^\beta_{f_1}\nn\\ 
&&-i(m_B+m_{f_1})V^{B\to f_1}_1(m^2_{K^*})(\epsilon_{f_1}\cdot \epsilon_{K^*}) 
+\frac{iV^{B\to f_1}_2(m^2_{K^*})}{(m_B+m_{f_1})}(\epsilon_{f_1} \cdot p_B)
(\epsilon_{K^*} \cdot p_B)\Bigg)\nn\\
&&+m_{f_1} f_{f_1} 
\Bigg\{ V_{ub}V^*_{us}\ a_{2}
-V_{tb}V^*_{ts}\left[2a_3-2a_5-\frac{1}{2}(a_7-a_9)\right]\Bigg\}
\Bigg(\frac{2V^{B\to K^*(m^2_{f_1})}}{(m_B+m_{K^*})}
\epsilon_{\mu\nu\alpha\beta}\epsilon^\mu_{f_1}\epsilon^\nu_{K^*}
p^\alpha_B p^\beta_{K^*}\nn\\
&&-i(m_B+m_{K^*})A^{B\to K^*}_1(m^2_{f_1})(\epsilon_{K^*}\cdot \epsilon_{f_1})
+\frac{iA^{B\to K^*}_2(m^2_{f_1})}{(m_B+m_{K^*})}(\epsilon_{K^*}\cdot p_B)
(\epsilon_{f_1} \cdot p_B)\Bigg)\ea

\ba {\cal M}(B^- \to f_1 K^{*-}) &=& -m_{K^*} f_{K^*}
\Bigg\{V_{ub}V^*_{us} a_1 -V_{tb}V^*_{ts}(a_4+a_{10})\Bigg\}
\Bigg(\frac{2A^{B\to f_1}(m^2_{K^*})}{(m_B+m_{f_1})}
\epsilon_{\mu\nu\alpha\beta}\epsilon^\mu_{K^*}\epsilon^\nu_{f_1}
p^\alpha_B p^\beta_{f_1}\nn\\ 
&&-i(m_B+m_{f_1})V^{B\to f_1}_1(m^2_{K^*})(\epsilon_{f_1}\cdot \epsilon_{K^*})+ 
\frac{iV^{B\to f_1}_2(m^2_{K^*})}{(m_B+m_{f_1})}(\epsilon_{f_1} \cdot p_B)
(\epsilon_{K^*}\cdot p_B)\Bigg)\nn\\
&&+m_{f_1} f_{f_1}
\Bigg\{V_{ub}V^*_{us}a_2- V_{tb}V^*_{ts}
\left[2a_3-2a_5-\frac{1}{2}(a_7-a_9)\right]\Bigg\}
\Bigg(\frac{2V^{B\to K^*}(m^2_{f_1})}{(m_B+m_{K^*})} 
\epsilon_{\mu\nu\alpha\beta}\epsilon^\mu_{f_1}\epsilon^\nu_{K^*}
p^\alpha_B p^\beta_{K^*}\nn\\
&&- i(m_B+m_{K^*})A^{B\to K^*}_1(m^2_{f_1})(\epsilon_{K^*}\cdot \epsilon_{f_1}) + 
\frac{iA^{B\to K^*}_2(m^2_{f_1})}{(m_B+m_{K^*})}(\epsilon_{K^*}\cdot p_B)
(\epsilon_{f_1}\cdot p_B)\Bigg)\ea

\ba {\cal M}(\bar B^0 \to f_1 \omega) &=& -m_\omega f_\omega
\Bigg\{ V_{ub}V^*_{ud}\ a_2 -V_{tb}V^*_{td}\left[2a_3+a_4+2a_5+\frac{1}{2}
(a_7+a_9-a_{10})\right]\Bigg\}
\Bigg(\frac{2A^{B\to f_1}(m^2_\omega)}{(m_B+m_{f_1})}
\epsilon_{\mu\nu\alpha\beta}\epsilon^\mu_\omega\epsilon^\nu_{f_1}
p^\alpha_B p^\beta_{f_1}\nn\\
&&-i(m_B+m_{f_1})V^{B\to f_1}_1(m^2_\omega)(\epsilon_{f_1} \cdot \epsilon_\omega) + 
\frac{iV^{B\to f_1}_2(m^2_\omega)}{(m_B+m_{f_1})}(\epsilon_{f_1}\cdot p_B)
(\epsilon_\omega \cdot p_B)\Bigg)\nn\\
&& +m_{f_1} f_{f_1} 
\Bigg\{ V_{ub}V^*_{ud}\ a_2-V_{tb}V^*_{td}\left[2a_3+a_4-2a_5-
\frac{1}{2}(a_7-a_9+a_{10})\right]\Bigg\}
\Bigg(\frac{2V^{B\to \omega}(m^2_{f_1})}{(m_B+m_\omega)} 
\epsilon_{\mu\nu\alpha\beta}\epsilon^\mu_{f_1}\epsilon^\nu_\omega
p^\alpha_B p^\beta_\omega\nn\\
&&-i(m_B+m_\omega)A^{B\to \omega}_1(m^2_{f_1})(\epsilon_\omega \cdot \epsilon_{f_1}) + 
\frac{iA^{B\to \omega}_2(m^2_{f_1})}{(m_B+m_\omega)}
(\epsilon_\omega \cdot p_B)(\epsilon_{f_1}\cdot p_B)\Bigg)\ea

\ba {\cal M}(\bar B^0 \to f_1 \phi) &=& m_\phi f_\phi
\Bigg\{V_{tb}V^*_{td} \left[a_3+a_5-\frac{1}{2}(a_7+a_9)\right]\Bigg\}
\Bigg(\frac{2A^{B\to f_1}(m^2_\phi)}{(m_B+m_{f_1})}
\epsilon_{\mu\nu\alpha\beta}\epsilon^\mu_\phi\epsilon^\nu_{f_1}
p^\alpha_B p^\beta_{f_1}\nn\\
&&-i(m_B+m_{f_1})V^{B\to f_1}_1(m^2_\phi)(\epsilon_{f_1}\cdot \epsilon_\phi) + 
\frac{iV^{B\to f_1}_2(m^2_\phi)}{(m_B+m_{f_1})}(\epsilon_{f_1} \cdot p_B)
(\epsilon_\phi \cdot p_B)\Bigg)\ea

\ba {\cal M}(\bar B^0 \to K_1\rho^+) &=& m_{K_1}f_{K_1} 
\Bigg\{ V_{ub}V^*_{us}\ a_1 -V_{tb}V^*_{ts}(a_4+a_{10})\Bigg\}
\Bigg(\frac{2V^{B\to \rho}(m^2_{K_1})}{(m_B+m_\rho)}
\epsilon_{\mu\nu\alpha\beta}\epsilon^\mu_{K_1}\epsilon^\nu_\rho
p^\alpha_B p^\beta_\rho\nn\\
&&-i(m_B+m_\rho)A^{B\to \rho}_1(m^2_{K_1})(\epsilon_\rho \cdot \epsilon_{K_1}) + 
\frac{iA^{B\to \rho}_2(m^2_{K_1})}{(m_B+m_\rho)}(\epsilon_\rho \cdot p_B)
(\epsilon_{K_1} \cdot p_B)\Bigg)\ea

\ba {\cal M}(\bar B^0\to \bar K^0_1\rho^0) &=& -m_\rho f_\rho 
\Bigg\{V_{ub}V^*_{us}\ a_2 - V_{tb}V^*_{ts} {3\over 2}(a_7 + a_9)\Bigg\}
\Bigg(\frac{2A^{B\to K_1}(m^2_\rho)}{(m_B+m_{K_1})}
\epsilon_{\mu\nu\alpha\beta}\epsilon^\mu_\rho\epsilon^\nu_{K_1}
p^\alpha_B p^\beta_{K_1}\nn\\
&&-i(m_B+m_{K_1})V^{B\to K_1}_1(m^2_\rho)(\epsilon_{K_1}\cdot \epsilon_\rho) +
\frac{iV^{B\to K_1}_2(m^2_\rho)}{(m_B+m_{K_1})}(\epsilon_{K_1}\cdot p_B)
(\epsilon_\rho \cdot p_B)\Bigg)\nn\\
&&- m_{K_1} f_{K_1}
\Bigg\{ V_{tb}V^*_{ts} \left(a_4-\frac{1}{2}a_{10}\right)\Bigg\}
\Bigg(\frac{2V^{B\to \rho}(m^2_{K_1})}{(m_B+m_\rho)}
\epsilon_{\mu\nu\alpha\beta}\epsilon^\mu_{K_1}\epsilon^\nu_\rho
p^\alpha_B p^\beta_\rho\nn\\
&&-i(m_B+m_\rho)A^{B\to \rho}_1(m^2_{K_1})(\epsilon_\rho^0\cdot \epsilon_{K_1}) +
\frac{iA^{B\to \rho}_2(m^2_{K_1})}{(m_B+m_\rho)}(\epsilon_\rho \cdot p_B)
(\epsilon_{K_1}\cdot p_B)\Bigg)\ea

\ba {\cal M}(B^-\to K^-_1\rho^0) &=& -m_\rho f_\rho
\Bigg\{V_{ub}V^*_{us} a_2 -V_{tb}V^*_{ts}\frac{3}{2}(a_7+a_9)\Bigg\}
\Bigg(\frac{2A^{B\to K_1}(m^2_\rho)}{(m_B+m_{K_1})}
\epsilon_{\mu\nu\alpha\beta}\epsilon^\mu_\rho\epsilon^\nu_{K_1}
p^\alpha_B p^\beta_{K_1}\nn\\
&&-i(m_B+m_{K_1})V^{B\to K_1}_1(m^2_\rho)(\epsilon_{K_1}\cdot \epsilon_\rho) + 
\frac{iV^{B\to K_1}_2(m^2_\rho)}{(m_B+m_{K_1})}(\epsilon_{K_1}\cdot p_B)
(\epsilon_\rho \cdot p_B)\Bigg)\nn\\
&&+m_{K_1} f_{K_1}
\Bigg\{V_{ub}V^*_{us} a_1 -V_{tb}V^*_{ts}(a_4+a_{10})\Bigg\}
\Bigg(\frac{2V^{B\to \rho}(m^2_{K_1})}{(m_B+m_\rho)}
\epsilon_{\mu\nu\alpha\beta}\epsilon^\mu_{K_1}\epsilon^\nu_\rho
p^\alpha_B p^\beta_\rho\nn\\
&&-i(m_B+m_\rho)A^{B\to \rho}_1(m_{K_1})(\epsilon_\rho\cdot \epsilon_{K_1}) + 
\frac{iA^{B\to \rho}_2(m^2_{K_1})}{(m_B+m_\rho)}(\epsilon_\rho \cdot p_B)
(\epsilon_{K_1}\cdot p_B)\Bigg)\ea

\ba {\cal M}(B^-\to \bar K^0_1\rho^-) &=& -m_{K_1} f_{K_1}
\Bigg\{ V_{tb}V^*_{ts}\left(a_4-\frac{1}{2}a_{10}\right)\Bigg\}
\Bigg(\frac{2V^{B\to \rho}(m^2_{K^0_1})}{(m_B+m_\rho)}
\epsilon_{\mu\nu\alpha\beta}\epsilon^\mu_{K_1}\epsilon^\nu_\rho
p^\alpha_B p^\beta_\rho\nn\\
&&-i(m_B+m_\rho)A^{B\to \rho}_1(m^2_{K_1})(\epsilon_\rho\cdot \epsilon_{K_1}) +
\frac{iA^{B\to \rho}_2(m^2_{K_1})}{(m_B+m_\rho)}
(\epsilon_\rho \cdot p_B)(\epsilon_{K_1} \cdot p_B)\Bigg)\ea

\ba {\cal M}(\bar B^0\to \bar K^0_1\omega) &=& -m_\omega f_\omega 
\Bigg\{ V_{ub}V^*_{us}\ a_2 -V_{tb}V^*_{ts}\left[2(a_3+a_5)+
\frac{1}{2}(a_7+a_9)\right]\Bigg\}
\Bigg(\frac{2A^{B\to K_1}(m^2_\omega)}{(m_B+m_{K_1})}
\epsilon_{\mu\nu\alpha\beta}\epsilon^\mu_\omega\epsilon^\nu_{K_1}
p^\alpha_B p^\beta_{K_1}\nn\\
&&-i(m_B+m_{K_1})V^{B\to K_1}_1(m^2_\omega)(\epsilon_{K_1}\cdot \epsilon_\omega) +
\frac{iV^{B\to K_1}_2(m^2_\omega)}{(m_B+m_{K_1})}(\epsilon_{K_1}\cdot p_B)
(\epsilon_\omega \cdot p_B)\Bigg)\nn\\
&&-m_{K_1} f_{K_1}
V_{tb}V^*_{ts}\Bigg\{\left(a_4-\frac{1}{2}a_{10}\right)\Bigg\}
\Bigg(\frac{2V^{B\to \omega}(m^2_{K_1})}{(m_B+m_\omega)}
\epsilon_{\mu\nu\alpha\beta}\epsilon^\mu_{K_1}\epsilon^\nu_\omega
p^\alpha_B p^\beta_\omega\nn\\
&&-i(m_B+m_\omega)A^{B\to \omega}_1(m^2_{K_1})(\epsilon_\omega \cdot \epsilon_{K_1}) +
\frac{iA^{B\to \omega}_2(m^2_{K_1})}{(m_B+m_\omega)}
(\epsilon_\omega \cdot p_B)(\epsilon_{K_1}\cdot p_B)\Bigg)\ea

\ba {\cal M}(B^-\to K^-_1\omega) &=& -m_\omega f_\omega
\Bigg\{V_{ub}V^*_{us} a_2 -V_{tb}V^*_{ts}
\left[2(a_3+a_5)+\frac{1}{2}(a_7+a_9)\right]\Bigg\}
\Bigg(\frac{2A^{B\to K_1}(m^2_\omega)}{(m_B+m_{K_1})}
\epsilon_{\mu\nu\alpha\beta}\epsilon^\mu_\omega\epsilon^\nu_{K_1}
p^\alpha_B p^\beta_{K_1}\nn\\
&&-i(m_B+m_{K_1})V^{B\to K_1}_1(m^2_\omega)
(\epsilon_{K_1} \cdot \epsilon_\omega)+\frac{iV^{B\to K_1}_2(m^2_\omega)}
{(m_B+m_{K_1})}(\epsilon_{K_1} \cdot p_B)(\epsilon_\omega \cdot p_B)\Bigg)\nn\\
&&+m_{K_1} f_{K_1} 
\Bigg\{V_{ub}V^*_{us}a_1-V_{tb}V^*_{ts}(a_4+a_{10})\Bigg\}
\Bigg(\frac{2V^{B\to \omega}(m^2_{K_1})}{(m_B+m_\omega)}
\epsilon_{\mu\nu\alpha\beta}\epsilon^\mu_{K_1}\epsilon^\nu_\omega
p^\alpha_B p^\beta_\omega\nn\\
&&-i(m_B+m_\omega)A^{B\to \omega}_1(m^2_{K_1})(\epsilon_\omega \cdot \epsilon_{K_1}) 
+ \frac{iA^{B\to \omega}_2(m^2_{K_1})}{(m_B+m_\omega)}(\epsilon_\omega \cdot p_B)
(\epsilon_{K_1}\cdot p_B)\Bigg)\ea

\ba {\cal M}(\bar B^0\to \bar K^0_1\phi) &=& m_\phi f_\phi
\Bigg\{V_{tb}V^*_{ts}\left[a_3+a_4+a_5-\frac{1}{2}(a_7+a_9+a_{10})\right]\Bigg\}
\Bigg(\frac{2A^{B\to K_1}(m^2_\phi)}{(m_B+m_{K_1})}
\epsilon_{\mu\nu\alpha\beta}\epsilon^\mu_\phi\epsilon^\nu_{K_1}
p^\alpha_B p^\beta_{K_1}\nn\\
&&-i(m_B+m_{K_1})V^{B\to K_1}_1(m^2_\phi)
(\epsilon_{K_1}\cdot \epsilon_\phi)+\frac{iV^{B\to K_1}_2(m^2_\phi)}
{(m_B+m_{K_1})}(\epsilon_{K_1}\cdot p_B)(\epsilon_\phi \cdot p_B)\Bigg)\ea

\ba {\cal M}(B^0\to K^0_1\bar K^{*0}) &=& m_{K^*}f_{K^*} 
\Bigg\{ V_{tb}V^*_{td} \left(a_4-{1\over 2}a_{10}\right)\Bigg\}
\Bigg(\frac{2A^{B\to K_1}(m^2_{K^*})}{(m_B+m_{K_1})}
\epsilon_{\mu\nu\alpha\beta}\epsilon^\mu_{K^*}\epsilon^\nu_{K_1}
p^\alpha_B p^\beta_{K_1}\nn\\
&&-i(m_B+m_{K_1})V^{B\to K_1}_1(m^2_{K^*})(\epsilon_{K_1}\cdot \epsilon_{K*}) + 
\frac{iV^{B\to K_1}_2(m^2_{K^*})}{(m_B+m_{K_1})}(\epsilon_{K_1}\cdot p_B)
(\epsilon_{K^*}\cdot p_B)\Bigg)\ea

\section{Matrix elements for $B$ decays to two axial mesons}

\ba {\cal M}(\bar B^0 \to a^-_1 a^+_1) &=& m_{a_1} f_{a_1}
\Bigg\{ V_{ub}V^*_{ud} a_1 - V_{tb}V^*_{td}(a_4+a_{10})\Bigg\}
\Bigg(\frac{2A^{B\to a_1}(m^2_{a_1})}{(m_B+m_{a_1})}
\epsilon_{\mu\nu\alpha\beta}\epsilon^\mu_{a^+_1}\epsilon^\nu_{a^-_1}
p^\alpha_B p^\beta_{a_1}\nn\\
&&-i(m_B+m_{a_1})V^{B\to a_1}_1(m^2_{a_1})(\epsilon_{a_1}\cdot \epsilon_{a_1}) + 
\frac{iV^{B\to a_1}_2(m^2_{a_1})}{(m_B+m_{a_1})}(\epsilon_{a_1}\cdot p_B)
(\epsilon_{a_1} \cdot p_B)\Bigg)\ea

\ba {\cal M}(\bar B^0 \to a^0_1 a^0_1) &=& m_{a_1} f_{a_1}
\Bigg\{ V_{ub}V^*_{ud} 2 a_2  -2 V_{tb}V^*_{td}\left[-a_4-{1\over
2}(3a_7-3a_9-a_{10})\right] \Bigg\}
\Bigg(\frac{2A^{B\to a_1}(m^2_{a_1})}{(m_B+m_{a_1})}
\epsilon_{\mu\nu\alpha\beta}\epsilon^\mu_{a_1}\epsilon^\nu_{a^0_1}
p^\alpha_B p^\beta_{a_1}\nn\\
&&-i(m_B+m_{a_1})V^{B\to a_1}_1(m^2_{a_1})(\epsilon_{a_1} \cdot \epsilon_{a_1}) + 
\frac{iV^{B\to a_1}_2(m^2_{a_1})}{(m_B+m_{a_1})}(\epsilon_{a_1}\cdot p_B)
(\epsilon_{a_1} \cdot p_B)\Bigg)\ea

\ba {\cal M}(B^- \to a^-_1 a^0_1) &=& m_{a_1} f_{a_1}
\Bigg\{ V_{ub}V^*_{ud} a_1 -V_{tb}V^*_{td}{3\over 2}(-a_7+a_9+a_{10})\Bigg\}
\Bigg(\frac{2A^{B\to a_1}(m^2_{a_1})}{(m_B+m_{a_1})}
\epsilon_{\mu\nu\alpha\beta}\epsilon^\mu_{a^-_1}\epsilon^\nu_{a^+_1}
p^\alpha_B p^\beta_{a_1}\nn\\ 
&&-i(m_B+m_{a_1})V^{B\to a_1}_1(m^2_{a_1})(\epsilon_{a_1} \cdot \epsilon_{a_1}) 
+\frac{iV^{B\to a_1}_2(m^2_{a_1})}{(m_B+m_{a_1})}(\epsilon_{a_1}\cdot p_B)
(\epsilon_{a_1} \cdot p_B)\Bigg)\nn\\
&&+m_{a_1} f_{a_1}
\Bigg\{ V_{ub}V^*_{ud} a_2 \Bigg\}
\Bigg(\frac{2A^{B\to a_1}(m^2_{a_1})}{(m_B+m_{a_1})} 
\epsilon_{\mu\nu\alpha\beta}\epsilon^\mu_{a^+_1}\epsilon^\nu_{a^-_1}
p^\alpha_B p^\beta_{a_1}\nn\\ 
&&-i(m_B+m_{a_1})V^{B\to a_1}_1(m^2_{a_1})(\epsilon_{a_1}\cdot \epsilon_{a_1}) + 
\frac{iV^{B\to a_1}_2(m^2_{a_1})}{(m_B+m_{a_1})}(\epsilon_{a_1} \cdot p_B)
(\epsilon_{a_1} \cdot p_B)\Bigg)\ea

\ba {\cal M}(\bar B^0 \to a^0_1 f_1) &=& m_{f_1} f_{f_1}
\Bigg\{ V_{ub}V^*_{ud} a_2 -V_{tb}V^*_{td}\left[2a_3+a_4-2a_5-
{1\over 2}(a_7-a_9+a_{10})\right]\Bigg\}
\Bigg(\frac{2A^{B\to a_1}(m^2_{f_1})}{(m_B+m_{a_1})}
\epsilon_{\mu\nu\alpha\beta}\epsilon^\mu_{f_1}\epsilon^\nu_{a_1}
p^\alpha_B p^\beta_{a_1}\nn\\
&&-i(m_B+m_{a_1})V^{B\to a_1}_1(m^2_{f_1})(\epsilon_{a_1} \cdot \epsilon_{f_1}) + 
\frac{iV^{B\to a_1}_2(m^2_{f_1})}{(m_B+m_{a_1})}(\epsilon_{a_1}\cdot p_B)
(\epsilon_{f_1} \cdot p_B)\Bigg)\nn\\
&&+ m_{a_1} f_{a_1}
\Bigg\{ V_{ub}V^*_{ud} a_2 -V_{tb}V^*_{td}\left[-a_4-
{3\over 2}(a_7-a_9)+{1\over 2}a_{10}\right] \Bigg\}
\Bigg(\frac{2A^{B\to f_1}(m^2_{a_1})}{(m_B+m_{f_1})} 
\epsilon_{\mu\nu\alpha\beta}\epsilon^\mu_{a_1}\epsilon^\nu_{f_1}
p^\alpha_B p^\beta_{f_1}\nn\\
&&-i(m_B+m_{f_1})V^{B\to f_1}_1(m^2_{a_1})(\epsilon_{f_1}\cdot \epsilon_{a_1})+ 
\frac{iV^{B\to f_1}_2(m^2_{a_1})}{(m_B+m_{f_1})}(\epsilon_{f_1} \cdot p_B)
(\epsilon_{a_1} \cdot p_B)\Bigg)\ea

\ba {\cal M}(B^- \to a^-_1 f_1) &=& m_{f_1} f_{f_1}
\Bigg\{ V_{ub}V^*_{ud}\ a_2 -V_{tb}V^*_{td}\left[2a_3+a_4-2a_5-{1\over 2}
(a_7-a_9+a_{10})\right]\Bigg\}
\Bigg(\frac{2A^{B\to a_1}(m^2_{f_1})}{(m_B+m_{a_1})}
\epsilon_{\mu\nu\alpha\beta}\epsilon^\mu_{f_1}\epsilon^\nu_{a_1}
p^\alpha_B p^\beta_{a_1}\nn\\
&&-i(m_B+m_{a_1})V^{B\to a_1}_1(m^2_{f_1})(\epsilon_{a_1}\cdot \epsilon_{f_1}) + 
\frac{iV^{B\to a_1}_2(m^2_{f_1})}{(m_B+m_{a_1})}(\epsilon_{a_1}\cdot p_B)
(\epsilon_{f_1} \cdot p_B)\Bigg)\nn\\
&& +m_{a_1} f_{a_1} 
\Bigg\{V_{ub}V^*_{ud}\ a_1 -V_{tb}V^*_{td}(a_4+a_{10})\Bigg\}
\Bigg(\frac{2A^{B\to f_1}(m^2_{a_1})}{(m_B+m_{f_1})} 
\epsilon_{\mu\nu\alpha\beta}\epsilon^\mu_{a_1}\epsilon^\nu_{f_1}
p^\alpha_B p^\beta_{f_1}\nn\\
&&-i(m_B+m_{f_1})V^{B\to f_1}_1(m^2_{a_1})(\epsilon_{f_1}\cdot \epsilon_{a_1}) + 
\frac{iV^{B\to f_1}_2(m^2_{a_1})}{(m_B+m_{f_1})}(\epsilon_{f_1} \cdot p_B)
(\epsilon_{a_1}\cdot p_B)\Bigg)\ea

\ba {\cal M}(\bar B^0 \to a^+_1 K^-_1) &=& m_{K_1}f_{K_1}
\Bigg\{ V_{ub}V^*_{us} a_1 -V_{tb}V^*_{ts}(a_4+a_{10})\Big\}
\Bigg(\frac{2A^{B\to a_1}(m^2_{K_1})}{(m_B+m_{a_1})} 
\epsilon_{\mu\nu\alpha\beta}\epsilon^\mu_{K_1}\epsilon^\nu_{a_1}
p^\alpha_B p^\beta_{a_1}\nn\\
&&-i(m_B+m_{a_1})V^{B\to a_1}_1(m^2_{K_1})(\epsilon_{a_1}\cdot 
\epsilon_{K_1}) + \frac{iV^{B\to a_1}_2(m^2_{K_1})}{(m_B+m_{a_1})}
(\epsilon_{a_1}\cdot p_B)(\epsilon_{K_1} \cdot p_B)\Bigg)\ea

\ba {\cal M}(\bar B^0\to a^0_1\bar K^0_1) &=& m_{a_1} f_{a_1}
\Bigg\{V_{ub}V^*_{us} a_2 - V_{tb}V^*_{ts}{3\over 2}(-a_7+a_9)\Bigg\}
\Bigg(\frac{2A^{B\to K_1}(m^2_{a_1})}{(m_B+m_{K_1})}
\epsilon_{\mu\nu\alpha\beta}\epsilon^\mu_{a_1}\epsilon^\nu_{K_1}
p^\alpha_B p^\beta_{K_1}\nn\\ 
&&-i(m_B+m_{K_1})V^{B\to K_1}_1(m^2_{a_1})(\epsilon_{K_1}\cdot \epsilon_{a_1}) +
\frac{iV^{B\to K_1}_2(m^2_{a_1})}{(m_B+m_{K_1})}(\epsilon_{K_1}\cdotp_B)
(\epsilon_{a_1}\cdot p_B)\Bigg)\nn\\
&&-m_{K_1}f_{K_1}
\Bigg\{V_{tb}V^*_{ts}\left(a_4-\frac{1}{2}a_{10}\right)\Bigg\}
\Bigg(\frac{2A^{B\to a_1}(m^2_{K_1})}{(m_B+m_{a_1})}
\epsilon_{\mu\nu\alpha\beta}\epsilon^\mu_{K_1}\epsilon^\nu_{a_1}
p^\alpha_B p^\beta_{a_1}\nn\\
&&-i(m_B+m_{a_1})V^{B\to a_1}_1(m^2_{K_1})
(\epsilon_{a_1}\cdot \epsilon_{K_1})+\frac{iV^{B\to a_1}_2(m^2_{K_1})}{(m_B+m_{a_1})}
(\epsilon_{a_1} \cdot p_B)(\epsilon_{K_1}\cdot p_B)\Bigg)\ea

\ba {\cal M}(B^-\to a^0_1 K^-_1) &=& m_{K_1}f_{K_1}
\Bigg\{V_{ub}V^*_{us}-V_{tb}V^*_{ts}(a_4+a_{10})\Bigg\}
\Bigg(\frac{2A^{B\to a_1}(m^2_{K_1})}{(m_B+m_{a_1})} 
\epsilon_{\mu\nu\alpha\beta}\epsilon^\mu_{K_1}\epsilon^\nu_{a_1}
p^\alpha_B p^\beta_{a_1}\nn\\
&&-i(m_B+m_{a_1})V^{B\to a_1}_1(m^2_{K_1})(\epsilon_{a_1}\cdot \epsilon_{K_1})+
\frac{iV^{B\to a_1}_2(m^2_{K_1})}{(m_B+m_{a_1})}(\epsilon_{a_1}\cdot p_B)
(\epsilon_{K_1}\cdot p_B)\Bigg)\nn\\
&&+m_{a_1} f_{a_1}
\Bigg\{ V_{ub} V^*_{us} + V_{tb}V^*_{ts}{3\over 2}(a_7-a_9)\Bigg\}
\Bigg(\frac{2A^{B\to K_1}(m^2_{a_1})}{(m_B+m_{K_1})}
\epsilon_{\mu\nu\alpha\beta}\epsilon^\mu_{a_1}\epsilon^\nu_{K_1}
p^\alpha_B p^\beta_{K_1}\nn\\
&&-i(m_B+m_{K_1})V^{B\to K_1}_1(m^2_{a_1})(\epsilon_{K_1} \cdot \epsilon_{a_1}) + 
\frac{iV^{B\to K_1}_2(m^2_{a_1})}{(m_B+m_{K_1})}(\epsilon_{K_1}\cdot p_B)
(\epsilon_{a_1}\cdot p_B)\Bigg)\ea

\ba {\cal M}(B^-\to a^-_0\bar K^0_1) &=& -m_{K_1} f_{K_1} 
\Bigg\{ V_{tb}V^*_{ts}\left(a_4-\frac{1}{2}a_{10}\right)\Bigg\}
\Bigg(\frac{2A^{B\to a_1}(m^2_{K_1})}{(m_B+m_{a_1})}
\epsilon_{\mu\nu\alpha\beta}\epsilon^\mu_{K_1}\epsilon^\nu_{a_1}
p^\alpha_B p^\beta_{a_1}\nn\\ 
&&-i(m_B+m_{a_1})V^{B\to a_1}_1(m^2_{K_1})(\epsilon_{a_1}\cdot \epsilon_{K_1}) +
\frac{iV^{B\to a_1}_2(m^2_{K_1})}{(m_B+m_{a_1})}
(\epsilon_{a_1}\cdot p_B)(\epsilon_{K_1}\cdot p_B)\Bigg)\ea

\ba {\cal M}(\bar B^0\to \bar K^0_1f_1) &=& m_{f_1} f_{f_1}
\Bigg\{V_{ub}V^*_{us} a_2 - V_{tb}V^*_{ts}
\left[2a_3-2a_5-\frac{1}{2}(a_7-a_9)\right]\Bigg\}
\Bigg(\frac{2A^{B\to K_1}(m^2_{f_1})}{(m_B+m_{K_1})}
\epsilon_{\mu\nu\alpha\beta}\epsilon^\mu_{f_1}\epsilon^\nu_{K_1}
p^\alpha_B p^\beta_{K_1}\nn\\
&&-i(m_B+m_{K_1})
V^{B\to K_1}_1(m^2_{f_1})(\epsilon_{K_1}\cdot \epsilon_{f_1}) + 
\frac{iV^{B\to K_1}_2(m^2_{f_1})}{(m_B+m_{K_1})}(\epsilon_{K_1}\cdot p_B)
(\epsilon_{f_1}\cdot p_B)\Bigg)\nn\\
&& -m_{K_1} f_{K_1}
\Bigg\{ V_{tb}V^*_{ts}\left(a_4-{1\over2}a_{10}\right)\Bigg\}
\Bigg(\frac{2A^{B\to f_1}(m^2_{K_1})}{(m_B+m_{f_1})}
\epsilon_{\mu\nu\alpha\beta}\epsilon^\mu_{K_1}\epsilon^\nu_{f_1}
p^\alpha_B p^\beta_{f_1}\nn\\ 
&&-i(m_B+m_{f_1})V^{B\to f_1}_1(m^2_{K_1})(\epsilon_{f_1} \cdot \epsilon_{K_1}) + 
\frac{iV^{B\to f_1}_2(m^2_{K^0_1})}{(m_B+m_{f_1})}(\epsilon_{f_1}\cdot p_B)
(\epsilon_{K_1}\cdot p_B)\Bigg)\ea

\ba {\cal M}(B^-\to K^-_1 f_1) &=& m_{K_1}f_{K_1}
\Bigg\{V_{ub}V^*_{us} a_1 - V_{tb}V^*_{ts}(a_4+a_{10})\Bigg\}
\Bigg(\frac{2A^{B\to f_1}(m^2_{K_1})}{(m_B+m_{f_1})} 
\epsilon_{\mu\nu\alpha\beta}\epsilon^\mu_{K_1}\epsilon^\nu_{f_1}
p^\alpha_B p^\beta_{f_1}\nn\\
&&- i(m_B+m_{f_1})V^{B\to f_1}_1(m^2_{K_1})(\epsilon_{f_1}\cdot \epsilon_{K_1}) + 
\frac{iV^{B\to f_1}_2(m^2_{K_1})}{(m_B+m_{f_1})}(\epsilon_{f_1}\cdot p_B)
(\epsilon_{K_1}\cdot p_B)\Bigg)\nn\\
&& +m_{f_1} f_{f_1}
\Bigg\{V_{ub}V^*_{us}a_2-V_{tb}V^*_{ts}
\left[2a_3-2a_5-{1\over 2}(a_7-a_9)\right]\Bigg\}
\Bigg(\frac{2A^{B\to K_1}(m^2_{f_1})}{(m_B+m_{K_1})} 
\epsilon_{\mu\nu\alpha\beta}\epsilon^\mu_{f_1}\epsilon^\nu_{K_1}
p^\alpha_B p^\beta_{K_1}\nn\\
&&-i(m_B+m_{K_1})V^{B\to K_1}_1(m^2_{f_1})(\epsilon_{K_1}\cdot \epsilon_{f_1}) + 
\frac{iV^{B\to K_1}_2(m^2_{f_1})}{(m_B+m_{K_1})}(\epsilon_{K_1}\cdot p_B)
(\epsilon_{f_1}\cdot p_B)\Bigg)\ea

\end{appendix}


\begin{thebibliography}{99}

\bibitem{exp}
B. Aubert et al. (BABAR Collaboration), Phys. Rev. Lett. {\bf 97}, 051802 (2006), 
arXiv:hep-ex/0603050; 
K. Abe et al. (Belle Collaboration), arXiv:hep-ex/0507096; 
H. Yang et al. (Belle Collaboration), Phys. Rev. Lett. 94, 111802 (2005), arXiv:hep-ex/0412039; 
K. Abe  et al. (Belle Collaboration), arXiv:hep-ex/0408138; 
B. Aubert et al. (BaBar Collaboration), Phys. Rev. Lett. 90, 242001 (2003); 
D. Besson et al. (CLEO Collaboration), Phys. Rev. D 68, 032002 (2003).

\bibitem{nardulli05}
G. Nardulli and T. N. Pham, Phys. Lett. B {\bf 623}, 65 (2005).

\bibitem{ali}
A. Ali, G. Kramer and C-D. Lu, Phys. Rev. D {\bf 58}, 094009 (1998).

\bibitem{chen99}
Y-H. Chen, H-Y. Cheng, B. Tseng and K-C. Yang, Phys. Rev. D {\bf 60}, 094014 (1999).

\bibitem{pdg2006}
Particle Data Group, W. M. Yao et al., J. Phys. G {\bf 33}, 1 (2006).

\bibitem{kave96}
A. C. Katoch and R. C. Verma, J. Phys. G: Nucl. Part . Phys. {\bf 22}, 1765 (1996).

\bibitem{isgw89}
N. Isgur, D. Scora, B. Grinstein and M. B. Wise, Phys. Rev. D {\bf 39}, 799(1989).

\bibitem{nardulli06}
V. Laporta, G. Nardulli and T. N. Pham, Phys. Rev. D {\bf 74}, 054035 (2006).

\bibitem{chen05}
C. H. Chen, C. Q. Geng, Y. K. Hsiao and Z-T. Wei, Phys. Rev. D {\bf 72}, 054011 (2005).

\bibitem{cheng03}
H. Y. Cheng, Phys. Rev. D {\bf 68}, 094005 (2003).

\bibitem{isgw95}
D. Scora and N. Isgur, Phys. Rev. D {\bf 52}, 2783 (1995).

\bibitem{msmodel}
D. Melikhov and B. Stech, Phys. Rev. D {\bf 62}, 014006 (2000).

\bibitem{cheng06}
H-Y. Cheng and C-K. Chua, Phys. Rev. D {\bf 74}, 034020 (2006).

\bibitem{lee06}
J. P. Lee, Phys. Rev. D {\bf 74}, 074001 (2006).

\bibitem{mixing}
See for example: H-Y. Cheng, Phys. Rev. D {\bf 67}, 094007 (2003); 
M. Suzuki, Phys. Rev. D {\bf 47}, 1252 (1993); 
N. Isgur and M. B.  Wise, Phys. Lett. B {\bf 232}, 113 (1989).

\bibitem{lili06}
D-M. Li and Z. Li, Eur. Phys. J. A {\bf 28}, 369 (2006).

\bibitem{gronau03}
M. Gronau, D. Pirjol and D. Wyler, Phys. Rev. Lett. {\bf 90}, 051801 (2003).

\bibitem{gronau06}
M. Gronau and J. Zupan, Phys. Rev. D {\bf 73}, 057502 (2006).

\bibitem{wsb}
M. Wirbel, B. Stech and M. Bauer, Z. Phys. C {\bf 29}, 637 (1985); 
M. Bauer and M. Wirbel, Z. Phys. {\bf C 42}, 671 (1989).

\bibitem{ball}
P. Ball and R. Zwicky, Phys. Rev. D {\bf 71}, 014015 (2005);
{\bf 71}, 014029 (2005).

\bibitem{buras96}
G. Buchalla, A. J. Buras, and M. E. Lautenbacher, 
Rev. Mod. Phys. {\bf 68}, 1125 (1996).

\bibitem{buras95}
A. J. Buras, Nucl. Phys. B {\bf 434}, 606 (1995).

\bibitem{buras98}
A. J. Buras, arXiv:hep-ph/9806471.

\bibitem{buras99}
A. J. Buras and L. Silvestrini, Nucl. Phys. B {\bf 548}, 293 (1999).

\bibitem{wolfenstein}
L. Wolfenstein, Phys. Rev. Lett. {\bf 51}, 1945 (1983).

\bibitem{buras94}
A. J. Buras, M. E. Lautenbacher and G. Ostermaier, 
Phys. Rev. D {\bf 50}, 3433 (1994).

\bibitem{ckmfitter}
A. Hocker at al., Eur. Phys. J. C {\bf 21}, 225 (2001);
J. Charles et al., Eur. Phys. J. C {\bf 41}, 1 (2005).

\bibitem{utfit}
M. Bona, et al., JHEP {\bf 507}, 28 (2005); JHEP {\bf 603}, 080 (2006).

\bibitem{fusaoku}
H. Fusaoka and Y. Koide, Phys. Rev. D {\bf 57}, 3986 (1998).

\bibitem{leutwyler}
H. Leutwyler, Nucl. Phys. B (Proc. Suppl.) {\bf 64}, 223 (1998).

\bibitem{feldmann}
T. Feldmann, P. Kroll and B. Stech, Phys. Rev. D {\bf 58}, 114006 (1998); 
Phys. Lett. B {\bf 449}, 339 (1999).

\bibitem{cheng98}
H. Y. Cheng and B. Tseng, Phys. Rev. D {\bf 58}, 094005 (1998).

\bibitem{gabriel}
J. L. Diaz-Cruz, G. Lopez Castro and J. H. Munoz,  
Phys. Rev. D {\bf 54}, 2388 (1996).

\end{thebibliography}
\end{document}